\documentclass[12pt,reqno]{amsart}
\usepackage{amssymb}
\usepackage{graphicx}
\usepackage{enumerate}
\usepackage{algorithmic,algorithm}
\usepackage{latexsym, amsmath, amscd, amssymb}
\usepackage[english]{babel}
\usepackage{fancyhdr}
\usepackage{bm}
\usepackage{mathrsfs}   %$\mathscr{A}$
\usepackage{xspace}
\usepackage{natbib}
\usepackage{setspace}
\usepackage{enumerate}
\usepackage{verbatim}
\usepackage[english]{babel}
\usepackage{epsfig}
\RequirePackage{eufrak}
\usepackage{algorithmic,algorithm}

\newtheorem{lemma}{Lemma}[section]
\newtheorem{theorem}[lemma]{Theorem}
\newtheorem{corollary}[lemma]{Corollary}
\newtheorem{proposition}[lemma]{Proposition}
\newtheorem{definition}[lemma]{Definition}

\newtheorem{algo}[lemma]{Algorithm}

\newtheorem{example}[lemma]{Example}
\newtheorem{remark}[lemma]{Remark}

\newcommand{\op}{\operatorname}

\DeclareMathOperator{\supp}{supp} %support of a polynomial
\DeclareMathOperator{\Mon}{Mon} %Monomials of a polynomial
\DeclareMathOperator{\Ter}{Ter}  %Terms of a polynomial
\DeclareMathOperator{\Lm}{Lm}
\DeclareMathOperator{\Lt}{Lt}
\DeclareMathOperator{\Lc}{Lc}
\DeclareMathOperator{\Span}{Span}
\begin{document}
\title[Border Bases for Polynomial Rings over Noetherian Rings]
{Border Bases for Polynomial Rings over Noetherian Rings}
\author[ Ambedkar Dukkipati, Nithish Pai and Maria Francis]{Ambedkar Dukkipati, Nithish Pai and Maria Francis}

\email{ad@csa.iisc.ernet.in\\ nithish.pai@conduent.com\\ mariaf@csa.iisc.ernet.in }
\address{Dept. of Computer Science \& Automation\\Indian Institute of Science, Bangalore - 560012}
% \large
% 
%======================Abstract=================
\maketitle

%======================Abstract=================
\begin{abstract}
  The theory of border bases for zero-dimensional ideals has attracted several researchers in symbolic computation due to their numerical stability and mathematical elegance. As shown in (Francis \& Dukkipati, \textit{J. Symb. Comp.}, 2014), one can extend the concept of border bases over Noetherian rings whenever the corresponding residue class ring is finitely generated and free. In this paper we address the following problem: Can the concept of border basis over Noetherian rings exists for ideals when the corresponding residue class rings are finitely generated but need not necessarily be free modules?
  %Crucial to deal with this problem is a general form of Macauley basis theorem over Noetherian rings in this case, which we present in this paper.
  We present a border division algorithm and prove the termination of the algorithm for a special class of border bases. We show the existence of such border bases over Noetherian rings and present some characterizations in this regard. We also show that certain reduced Gr\"{o}bner bases over Noetherian rings are contained in this class of border bases. 
%  We develop a theory of border bases for ideals, $\mathfrak{a}$ in
%  polynomial rings over a Noetherian commutative ring, $A$ when the
%  residue class ring modulo $\mathfrak{a}$,
%  $A[x_{1},\ldots,x_{n}]/\mathfrak{a}$ is finitely generated but not
%  necessarily a free $A$-module. 
  %To develop this theory, we provide some general
  %forms of Macaulay basis theorems in this case. 

\end{abstract}
%\clearpage\maketitle
%\thispagestyle{empty}
%\keywords{Border Bases, Polynomial Rings over Noetherian Rings}
%==================================================================
\section{Introduction}
\noindent
 Gr\"obner bases theory for polynomial rings over fields  
gives an algorithmic technique to 
determine a vector space basis of the residue class ring modulo a
zero dimensional ideal \citep{Buchberger:1965:AnAlgorithmForFindingAbasis}. 
%We refer to this as  Macaulay-Buchberger basis
%theorem, as Buchberger's result is based on the work
%by~\cite{Macaulay:1916:AlgorithmicAlgebra}.   
The theory of Gr\"obner bases was extended to
polynomial rings over Noetherian commutative rings with unity a few
decades ago \citep[e.g.][]{Trinks:1978:grobnerforrings1,  
  Moller:1988:OnTheConstructionOfGBusingSyzygies,
  Zacharias:1978:GeneralizedGBinCommutativePolynomialRings}. 
  %but the Macaulay-Buchberger basis theorem was extended to 
%residue class polynomial rings over Noetherian commutative rings only recently in
%\citep{FrancisDukkipati:2014:OnReducedGrobnerBasisAndMBtheorem}.  The result was specific to free modules and uses Gr\"{o}bner bases to test algorithmically whether for a given monomial order, the residue class 
%polynomial ring over a  ring, $A$, has a free $A$-module representation or not. This result is also an elegant 
%generalization of the fact that $Z[x]/\langle f \rangle$ is free if 
%and only $f$ is monic, to a multivariate case over any Noetherian 
%ring. 
%In this paper we attempt to answer the following question: Can the Macaulay-Buchberger basis theorem \citep{Buchberger:1965:AnAlgorithmForFindingAbasis} 
%be generalized to residue class polynomial rings over Noetherian commutative rings with torsion? 
%$A[x_{1},\ldots,x_{n}]/ \mathfrak{a}$ that need not necessarily have a free
%$A$-module representation and study the possibility of generalizing Macaulay-Buchberger
%basis theorem to this case.
 %The generalization we propose in this paper helps us to develop a
%theory of border bases for polynomial rings over rings.  
%Even though various approaches were proposed for extending Gr\"obner basis theory to polynomial rings over rings \citep{kandri:1988:computing, Kapur:2009:zerodivisors}, these techniques only looked at extending basic definitions and concepts and validity of many important results were not explored \citep{Greuel:2011:TheGrobnerBasisOfTheIdealOfVanishingPolynomials}.
For a good exposition on Gr\"obner bases over rings one can refer to
\citep{AdamsLoustaunau:1994:AnIntroductionToGrobnerBases}.

Certain recent  developments in cryptography and other fields 
 have led to renewed interest in polynomial rings over
rings~\citep{Greuel:2011:TheGrobnerBasisOfTheIdealOfVanishingPolynomials}. For
instance, free  residue class rings over $\mathbb{Z}[x]$ called ideal
lattices~\citep{Lyubashevsky:2006:Ideallatticefirstdef} have been  
 shown to be isomorphic to integer lattices, an important
 cryptographic primitive~\citep{Ajtai:1996:Znascryptoprimitive} and
 certain cyclic lattices in $\mathbb{Z}[x]$ have been used in NTRU
 cryptographic schemes \citep{Hoffstein:1998:NTRUconf}. Boolean
 polynomial rings over a boolean ring is another important example of
 a polynomial ring over a ring since it can be used to solve
 combinatorial puzzles like Sudoku
 \citep{Sato:2011:BooleanGB}. Another example is the polynomial rings
 over $\mathbb{Z}/2^k$. They are used to prove the correctness of data
 paths in system-on-chip design
 \citep{Greuel:2011:TheGrobnerBasisOfTheIdealOfVanishingPolynomials}.   

%In the case of rings, the freeness of the residue class polynomial
%ring cannot be guaranteed.  
% The necessary and sufficient condition for determining the freeness
% of the  
%$A$-module is given in
% \citep{FrancisDukkipati:2014:OnReducedGrobnerBasisAndMBtheorem}.  
%The same paper also gives an algorithmic technique to determine an
%$A$-module basis of a free and finitely generated residue class
%polynomial ring. In this paper, we relax the condition of freeness and
%generalize the Macaulay-Buchberger theorem for residue class rings
%that are finitely generated, but need not be free.  The algorithmic
%technique we propose determines a generating set for
%$A[x_1,\ldots,x_n]/\mathfrak{a}$ using Gr\"obner bases  
%that satisfies a weaker form of the linear independence
%property. Using this generating set we extend the concept of border
%bases to polynomial rings over rings.  
Border bases, an alternative to Gr\"{o}bner bases, are well studied 
for polynomial rings over fields~\citep{KehreinKreuzerRobbiano:2005:AnAlgerbaistsViewOnBorderBases,KehreinKreuzer:2005:CharacterizationsOfBorderBases}.
%For
%a good exposition the reader can refer to %\citep{KreuzerRobbiano:2005:ComputationalCommutativeAlgebra2}.   
 Though border bases are restricted to zero dimensional ideals, 
the motivation for border bases  comes from the
 numerical stability of border bases over Gr\"{o}bner bases
 \citep{Stetter:2004:NumericalPolynomialAlgebra}. There has been
 considerable interest in the theory of border bases, from
 characterization~\citep{KehreinKreuzer:2005:CharacterizationsOfBorderBases} 
 to methods of
 computation~\citep{KehreinKreuzer:2006:ComputingBorderBases} to
 computational
 hardness~\citep{AnanthDukkipati:2012:ComplexityOfGrobnerBasisAndBorderBasisDetection}. 
 The concept of border bases can be easily extended to polynomial rings over rings 
 if the corresponding residue class ring has a free $A$-module representation w.r.t. some monomial order and is finitely
 generated as an
 $A$-module~\citep{FrancisDukkipati:2014:OnReducedGrobnerBasisAndMBtheorem}. In
 this paper, we study border bases for ideals in polynomial rings over
 Noetherian commutative rings in a more general set up.  
\section{Background \& Preliminaries}
\label{Preliminaries}
\subsection{Notations}
\noindent
%Let $\mathbb{N}$ denote the set of natural numbers including
%zero, and $\mathbb{Z}$ denote the ring of integers. 
A polynomial ring in 
indeterminates $x_{1},\ldots,x_{n}$ over a Noetherian, commutative ring $A$
is denoted by $A[x_{1},\ldots,x_{n}]$. Throughout this paper,
the rings we study are rings with unity. The set of all ideals of
$A$ is denoted by $\op{Id}(A)$.
When $A=\Bbbk$, where $\Bbbk$
is a field, we write it as $\Bbbk[x_{1},\ldots,x_{n}]$. 
%A monomial $x_{1}^{\alpha_{1}}\cdots x_{n}^{\alpha_{n}} $ is denoted
%by $x^{\alpha}$, 
%with the understanding that $x = (x_{1},\ldots,x_{n})$ and
%$\alpha = (\alpha_{1},\ldots,\alpha_{n}) \in {\mathbb{M}}^n$.
A monomial in indeterminates $x_{1},\ldots,x_{n}$ is denoted by
$x^{\alpha}$, where $\alpha \in \mathbb{N}^n$,
%The monoid isomorphism between the set of all monomials in indeterminates $x_1,\ldots, x_n$ and  
%${\mathbb{N}}^n$ allows us to denote the set of all monomials by ${\mathbb{N}}^n$.
and the set of all monomials is denoted by $\mathbb{M}^{n}$.
%We use $\mathbb{N}^{n}_{M}$
%and $\mathbb{N}^{n}_{<M}$ to represent the set of monomials of degree
%$M$ and monomials of degree less than $M$, respectively.
%When we need
%to deal with more than one monomial, in variables
%$x_{1},\ldots,x_{n}$, we index these monomials as
%$x^{\alpha_{1}},x^{\alpha_{2}},\ldots,x^{\alpha_{\ell}}$.
By
`term' we mean $cx^{\alpha}$, where $c \in A$ and $c \neq 0$. We will denote all the terms 
in $A[x_{1},\ldots,x_{n}]$ by $\mathbb{T}^{n}$.
%and all the monomials in 
% $A[x_{1},\ldots,x_{n}]$ by $\mathrm{Mon}(A[x_{1},\ldots,x_{n}])$. 
Let $T \subseteq \mathbb{T}^{n}$ be a set of terms, possibly
infinite. 
We define the monomial part of $T$, denoted by $\Mon(T)$, as
     $\Mon(T) = \{ x^{\alpha} \in \mathbb{M}^{n} : ax^{\alpha} \in T, \text{ for some nonzero } a \in A\}$.

A polynomial $f$ is denoted by
%\begin{displaymath}
   $f = \sum_{\alpha \in \Lambda} a_{\alpha}
   x^{\alpha}$,
%\end{displaymath}
where $a_{\alpha} \in A$, $\alpha \in \mathbb{N}^{n}$
and 
$\Lambda \subseteq {\mathbb{N}}^{n}$ is a finite set. $\Lambda$
is called the support of the polynomial $f$, denoted by $\supp(f)$. 
The set of monomials appearing with nonzero coefficients in $f$ is denoted
by $\Mon(f)$. The set of all terms appearing in $f$ is denoted by
$\Ter(f)$, i.e. $\Ter(f) = \{ a_{\alpha}x^{\alpha} \mid \alpha \in
\Lambda \}$. If $F$ is a set of
polynomials then $\Mon(F) = \bigcup_{f \in F} \Mon(f) $. Similarly,
$\Ter(F) = \bigcup_{f \in F} \Ter(f)$. Given a set of terms $T$
and an ideal $\mathfrak{a}$ in $A[x_{1},\ldots,x_{n}]$, the set of
residue class elements of $T$ modulo $\mathfrak{a}$ is denoted by
$T+\mathfrak{a}$. That is $T+ \mathfrak{a} = \{ ax^{\alpha} +
\mathfrak{a} : ax^{\alpha} \in T\}$. Given a set of polynomials $F$ the span of $F$ over $A$ is given   by
$\Span_A(F) = \{\sum_{i=1}^s a_if_i : a_i \in A, f_i \in F, s \in
\mathbb{N}\}$.  

With respect to a monomial order $\prec$, the leading monomial,
leading coefficient and leading term of a polynomial $f \in
A[x_{1},\ldots,x_{n}]$ are denoted by $\Lm(f)$, $\Lc(f)$ and $\Lt(f)$
respectively. That is, we have $\Lt(f) = \Lc(f)\Lm(f)$. Given a set of polynomials
$S$, possibly infinite, $\Lt(S)$ denotes the set of all leading
terms of polynomials in $S$.  
The leading term ideal (or initial ideal) of $S$ is denoted by
$\langle \Lt(S) \rangle$  and is given by $\langle \Lt(S) \rangle =
\langle  \{\mathrm{Lt}_\prec(f) \mid f \in S \} \rangle$. Similarly, the 
leading monomial ideal and the leading coefficient ideal of $S$ are denoted by
$\langle \Lm(S)\rangle$ and $\langle \Lc(S)\rangle$ respectively. Note that the leading coefficient ideal is an ideal in 
the coefficient ring, $A$.
The ideal generated by the polynomials $f_{1},\ldots, f_{s}$ is denoted by $\langle f_{1},\ldots, f_{s}
\rangle$. Also a set of polynomials $S$ is said to be monic w.r.t. a monomial order if the leading coefficient of each polynomial in $S$ is $1$. 
\subsection{Border bases over a field}
\noindent
Here we briefly recall definitions related to border bases. 
\begin{definition} 
  \label{Definition:OrderIdeal_Field}
A finite set of monomials $\mathcal{O} \subseteq {\mathbb{M}}^n$ is
said to be an order ideal if it is closed under forming divisors i.e., for
$x^{\alpha} \in {\mathbb{M}}^n$, if $x^{\beta} \in \mathcal{O}$ and $x^{\alpha} | x^{\beta}$, then $x^{\alpha} \in \mathcal{O}$. 
\end{definition}

\begin{definition}
Let $\mathcal{O}$ be an order ideal. The
border of $\mathcal{O}$ is the set $\partial \mathcal{O} =
%({\mathbb{N}}_{1}^n\mathcal{O}) \setminus \mathcal{O} =
(x_{1}\mathcal{O} \cup \ldots \cup x_{n}\mathcal{O}) \setminus
\mathcal{O}$. The first border closure of $\mathcal{O}$ is defined as
the set $\mathcal{O} \cup \partial \mathcal{O}$ and it is denoted by
$\overline{\partial \mathcal{O}}$.
\end{definition}
Note that $\overline{\partial \mathcal{O}}$ is also an order ideal. By convention, if  $\mathcal{O} = \emptyset$, then we set $\partial 
\mathcal{O} = \{1\}$. 

\begin{definition}
Let $\mathcal{O} = \{x^{\alpha_{1}}, \ldots, x^{\alpha_{s}}\}$ be an order ideal, and let $\partial
\mathcal{O} = \{x^{\beta_{1}}, \ldots, x^{\beta_{t}}\}$ be its
border. A set of polynomials $\mathcal{B} = \{b_{1}, \ldots, b_{t}\} \subseteq
\Bbbk[x_{1},\ldots, x_{n}]$ is called an $\mathcal{O}$-border prebasis
if the polynomials have the form
\begin{displaymath}
b_{j} = x^{\beta_{j}} - \sum_{i=1}^{s} c_{ij}x^{\alpha_{i}}\enspace,
\end{displaymath}
where $c_{ij} \in \Bbbk\ \mathrm{for}\ 1 \leq i \leq s$ and $1 \leq j
\leq t$.  
\end{definition}

\begin{definition}
Let $\mathcal{O} = \{x^{\alpha_{1}}, \ldots, x^{\alpha_{s}}\}$  be an order ideal and $\mathcal{B} = \{b_{1}, \ldots,
b_{t}\}$ be an $\mathcal{O}$-border prebasis consisting of polynomials
in $\mathfrak{a} \subseteq \Bbbk[x_{1},\ldots, x_{n}]$. We say that
the set $\mathcal{B}$ is an $\mathcal{O}$-border basis of $\mathfrak{a}$ if the
residue classes of $x^{\alpha_{1}}, \ldots, x^{\alpha_{s}}$ form a
$\Bbbk$-vector space basis of $\Bbbk[x_{1},\ldots,x_{n}]/\mathfrak{a}$.   
\end{definition}

The existence and uniqueness of border bases are established in~\citep{KehreinKreuzer:2005:CharacterizationsOfBorderBases}.
% Existence and Uniqueness of border bases
%\begin{theorem}\label{Unique Border and reduced Grobner bases}
%Let $\mathcal{O} = \{x^{\alpha_{1}}, \ldots, x^{\alpha_{s}}\}$ be an order ideal, let $\mathfrak{a}$ be a
%zero-dimensional ideal in $\Bbbk[x_{1},\ldots,x_{n}]$, and assume that the residue classes of the elements in $\mathcal{O}$ form a $\Bbbk$-vector
%space basis of $\Bbbk[x_{1},\ldots,x_{n}]/\mathfrak{a}$:
%\begin{enumerate}
% \item There exists a unique $\mathcal{O}$-border basis $H$ of $\mathfrak{a}$.
% \item Let $H'$ be an $\mathcal{O}$-border prebasis whose elements are in $\mathfrak{a}$. Then $G'$ is the $\mathcal{O}$-border basis of
% $\mathfrak{a}$.
% \item Let $\Bbbk$ be the field of definition of $\mathfrak{a}$. Then we have $H \subset %\Bbbk[x_{1},\ldots,x_{n}]$.
%\end{enumerate}
%\end{theorem}
Given an ideal one can show that there exists a border bases
that  do not correspond to Gr\"{o}bner bases for any term ordering. 
An example of such a border basis is given in \citep{KehreinKreuzer:2006:ComputingBorderBases}.
%==================================================================
\subsection{The $A$-module $A[x_{1},\ldots,x_{n}]/\mathfrak{a}$}
\noindent
\label{Theorem:MBBasisFreeCase}
%In this paper, we study border bases for ideals in polynomial rings
%where coefficients come from a Noetherian, commutative ring, $A$.
%In this section we recall some results
%from~\citep{FrancisDukkipati:2014:OnReducedGrobnerBasisAndMBtheorem}, 
%where Macaulay-Buchberger basis theorem for the case when residue
%class ring is a free $A$-module is presented. 
From now on, unless otherwise specified we deal with polynomials over a Noetherian, commutative ring $A$. 
%we look at polynomial rings where the coefficients come from Noetherian, commutative rings. 
%Unlike in polynomial rings over fields, here, the distinction between monomials and terms is more %marked.
%Consider a Noetherian, commutative ring, $A$  and an ideal $\mathfrak{a} \subseteq A[x_1,\ldots,x_n]$.
Given an ideal $\mathfrak{a}$ in $A[x_1,\ldots,x_n]$, using Gr\"{o}bner basis
methods one can give an $A$-module representation of residue class ring
$A[x_{1},\ldots,x_{n}]/\mathfrak{a}$ if it is finitely generated~\citep{FrancisDukkipati:2014:OnReducedGrobnerBasisAndMBtheorem}. We describe this briefly
below and for more details one can refer to~\citep{FrancisDukkipati:2014:OnReducedGrobnerBasisAndMBtheorem}. The notation is mostly borrowed from~\citep{AdamsLoustaunau:1994:AnIntroductionToGrobnerBases}.

Let $G = \{g_i: i = 1, \ldots, t\}$  be a Gr\"obner basis for $\mathfrak{a}$. 
For each monomial, $x^\alpha$, let $J_{x^{\alpha}} = \{i : \mathrm{lm}(g_i)\mid x^{\alpha},  g_i \in G \}$  
and $I_{J_{x^{\alpha}}} = \langle \{\mathrm{lc}(g_i) : i \in
J_{x^{\alpha}}\} \rangle$. Note that $I_{J_{x^{\alpha}}}$ is an ideal in $A$. 
We refer to $I_{J_{x^{\alpha}}}$ as the leading coefficient ideal
w.r.t. $G$.
Let $C_{J_{x^{\alpha}}}$ represent a set of coset representatives of the equivalence classes in  $A/I_{J_{x^{\alpha}}}$.
Given a polynomial, $f \in  A[x_1,\ldots,x_n]$, let $f =
\sum\limits_{i=1}^m a_i x^{\alpha_i}\hspace{2pt} \mathrm{mod} \hspace{2pt}\langle G \rangle$,
where $a_i \in A, i=1,\ldots,m$.   
% Consider the set $M = \{a_ix^{\alpha_i} : a_i \in C_{J_{x^{\alpha_i}}},  x^{\alpha_i} \in F\} $. 
If $A[x_{1},\ldots,x_{n}]/\langle G \rangle$ is an $A$-module generated by $m$ elements,
then corresponding to the coset representatives, $C_{J_{x^{\alpha_1}}}, \ldots, C_{J_{x^{\alpha_m}}}$, 
there exists an $A$-module isomorphism, 
\begin{equation} \label{equation}
\begin{split}
 \phi :  A[x_{1},\ldots,x_{n}]/\langle G \rangle &\longrightarrow A/I_{J_{x^{\alpha_1}}} \times \cdots \times A/I_{J_{x^{\alpha_m}}}\\
         \sum\limits_{i=1}^m a_i x^{\alpha_i} + \langle G \rangle &\longmapsto (c_1 +I_{J_{x^{\alpha_1}}}  , \cdots, c_m + I_{J_{x^{\alpha_m}}}) ,
\end{split}
\end{equation}
where $c_i = a_i \text{  mod  } I_{J_{x^{\alpha_i}}}$ and  $c_i \in C_{J_{x^{\alpha_i}}}$. 
We refer to $A/I_{J_{x^{\alpha_1}}} \times \cdots \times A/I_{J_{x^{\alpha_m}}}$ as the $A$-module representation of 
$A[x_{1},\ldots,x_{n}]/\mathfrak{a}$ w.r.t. $G$.  
If $I_{J_{x^{\alpha_i}}} = \{0\} $, we have $C_{J_{x^{\alpha_i}}} = A$,
$\text{for all } i = 1,\ldots,m$.  
This implies  $A[x_{1},\ldots,x_{n}]/\mathfrak{a} \cong A^m$, i.e. $A[x_{1},\ldots,x_{n}]/\mathfrak{a}$ has an $A$-module basis and it is free. In this case, we define $A[x_{1},\ldots,x_{n}]/\mathfrak{a} $ to have a ``free $A$-module representation w.r.t. $G$''. 
The necessary and sufficient condition for an $A$-module
$A[x_1,\ldots,x_n]/ \mathfrak{a} $ to have a free $A$-module representation is given in \citep{FrancisDukkipati:2014:OnReducedGrobnerBasisAndMBtheorem}. It makes use of the  the concept of 
`short reduced Gr\"obner basis' introduced in  \citep{FrancisDukkipati:2014:OnReducedGrobnerBasisAndMBtheorem} which we briefly describe below.
\begin{definition}
Let $\mathfrak{a}$ be an ideal in $A[x_1, \ldots, x_n]$ and let $G$ be its  reduced Gr\"obner basis as described in \citep{Pauer:2007:GrobnerBasesWithCoefficientsInRings}.
Consider the isomorphism in \eqref{equation}. $G$ is called a short reduced Gr\"obner basis if the size of the generating set of the leading coefficient ideal, $I_{J_{x^{\alpha}}}$, of each leading monomial, 
$x^\alpha$ in $G$, is minimal. 
\end{definition}
One can define reduced Gr\"obner bases over rings exactly as that of fields but it may not exist in all the cases. 
The definition of reduced Gr\"obner basis  given by \citep{Pauer:2007:GrobnerBasesWithCoefficientsInRings} is a generalization of the concept over fields to rings that also ensures the existence of such a basis for every ideal in the polynomial ring. 
%\begin{theorem}\label{necessarycondition}
%Let $\mathfrak{a} \subseteq A[x_1,\ldots,x_n]$ be a nonzero ideal and let $G$ be a short reduced Gr\"obner basis of $\mathfrak{a}$. 
%If $A[x_1,\ldots,x_n]/\mathfrak{a}$ has a free $A$-module representation w.r.t. $G$, then $G$ is monic. 
%\end{theorem}
%\begin{example}
% Let $G = \{3x,5x,y\}$ be the Gr\"obner basis of an ideal  $\mathfrak{a} $ in $\mathbb{Z}[x,y]$. 
% The short  reduced Gr\"obner basis of  $\mathfrak{a}$ is given by $G_\mathrm{red} = \{x,y\}$.
%It is monic and a $\mathbb{Z}$-module basis of  $\mathbb{Z}[x,y]/\mathfrak{a}$ is given by $\{1+\mathfrak{a}\}$. 
%\end{example}
The short reduced Gr\"obner basis of an ideal is not to be confused with strong Gr\"obner basis \citep[Definition 4.5.6.]{AdamsLoustaunau:1994:AnIntroductionToGrobnerBases}. Strong Gr\"obner basis exists only if the coefficient ring is a $\mathrm{PID}$. 
In a $\mathrm{PID}$, strong Gr\"obner basis coincides with the short reduced Gr\"obner basis. 
\begin{proposition}\citep[Proposition 3.12]{FrancisDukkipati:2014:OnReducedGrobnerBasisAndMBtheorem}\label{Proposition for characterization}
Let $\mathfrak{a} \subseteq A[x_1,\ldots,x_n]$ be a nonzero ideal such that $A[x_1,\ldots,x_n]/\mathfrak{a}$ is finitely generated. 
Let $G$ be a  short reduced Gr\"obner basis for $\mathfrak{a}$ w.r.t. some monomial ordering, $\prec$. Then, $A[x_1,\ldots,x_n]/\mathfrak{a}$ has a free $A$-module representation w.r.t. $G$
% \begin{displaymath}
% A[x_1,\ldots,x_n]/\mathfrak{a} \cong A^N, \hspace{10pt} \text{ for some } N \in \mathbb{N}
% \end{displaymath}
if and only if $G$ is monic.
\end{proposition}
%An example for this proposition can be found in Appendix~\ref{appendix_label}.
\begin{example}
 Let $G = \{3x,5x,y\}$ be the Gr\"obner basis of an ideal  $\mathfrak{a} $ in $\mathbb{Z}[x,y]$. 
 The short  reduced Gr\"obner basis of  $\mathfrak{a}$ is given by $G_\mathrm{red} = \{x,y\}$.
It is monic and a $\mathbb{Z}$-module basis of  $\mathbb{Z}[x,y]/\mathfrak{a}$ is given by $\{1+\mathfrak{a}\}$. 
\end{example}
With the
above result the concept of border bases can be extended to ideals in polynomial rings over
rings, in the cases where the corresponding residue class rings are
finitely generated and have a free $A$-module representation w.r.t. some monomial order \citep[Section 6]{FrancisDukkipati:2014:OnReducedGrobnerBasisAndMBtheorem}. One can show that all the
characterizations in
\citep{KehreinKreuzer:2005:CharacterizationsOfBorderBases} hold true
when the residue class ring is  free. For the sake of completeness, we
state the definition of border bases in this case below. 
\begin{definition}
Let  $\mathcal{O} =
\{x^{\alpha_1}, \ldots, x^{\alpha_s}\}$ be an order ideal and
$\mathcal{B} = \{b_1, \ldots, b_t\} \subseteq A[x_1,\ldots,x_n]$ be an
$\mathcal{O}$-border prebasis.  Let $\mathfrak{a}\subseteq
A[x_1,\ldots,x_n]$ be an ideal such that 
$A[x_1,\ldots,x_n]/\mathfrak{a}$ is finitely generated and is a free
$A$-module. Then $\mathcal{B}$ is said to be an $\mathcal{O}$-border
basis if  $ \mathcal{B}\subseteq \mathfrak{a}$ and $\mathcal{O}$ forms
an $A$-module basis for $A[x_1,\ldots,x_n]/\mathfrak{a}$.  
\end{definition}
\section{Order Functions and Border Prebasis Division Algorithm}
\label{Order Ideals and Border Prebasis Division Algorithm}
%We associate an ideal, $\mathcal{I}_{x^{\alpha}} \subsetneq A$ for a monomial
%$x^{\alpha} \in \mathbb{T}^{n}$.
%Let $\op{Id}(A)$ denote the set of all proper ideals in $A$. We define
%order ideal in this case  with respect to a finite subset of
%$\op{Id}(A) \times \mathbb{T}^{n}$.  

%We call the tuple, $(\mathcal{I}_{x^{\alpha}},x^{\alpha})$ formed by $\mathcal{I}_{x^{\alpha}}$ and the corresponding monomial, $x^{\alpha}$, an order tuple.
%We then define a set of order tuples as
%\begin{displaymath}
%B=\{(\mathcal{I}_{x^{\alpha}},x^{\alpha})\ |\ \mathcal{I}_{x^{\alpha}}\ \text{is an ideal
%  in}\ A \text{ such that}\ \mathcal{I}_{x^{\alpha}} \neq
%A\ \text{and}\ x^{\alpha} \in \mathbb{T}^{n}\}.
%\end{displaymath}
%Given an ideal $\mathfrak{a}$ in $A[x_{1},\ldots,x_{n}]/\mathfrak{a}$, in the case
%when it does not have a free $A$-module representation. That is, the leading coefficients of 
%polynomials in the short reduced Gr\"{o}bner basis need not be monic.
%Through out this section we assume a monomial order $\prec$.
\noindent
The following notion that we introduce in this paper is crucial to
the theory of border bases that we develop here. 
\begin{definition}[Order Function]
Let $\op{Id}(A)$ be the set of all ideals in the ring, $A$. 
A mapping $\mathcal{I}: \mathbb{M}^n
\rightarrow \op{Id}(A)$ is said to be an order function if $x^\beta \mid x^\alpha$ implies
$\mathcal{I}({x^\beta}) \subseteq \mathcal{I}({x^\alpha})$, for all
$x^\alpha, x^\beta \in \mathbb{M}^{n}$. From now on we denote
$\mathcal{I}({x^\alpha})$ by $\mathcal{I}_{x^\alpha}$.
\end{definition}
Clearly, the Gr\"{o}bner basis of an ideal `fixes' an order function. Consider the leading coefficient ideal, $I_{J_{x^{\alpha}}}$
that we constructed in Section~\ref{Theorem:MBBasisFreeCase}  
w.r.t. $G$. Since $J_{x^{\alpha}}$ is a saturated set, the mapping
$x^{\alpha} \mapsto I_{J_{x^{\alpha}}}$ is an order function, which is denoted by $\mathcal{I}^{(G)}$.      
\begin{definition}\label{coeffienctideal}
An order function $\mathcal{I}: \mathbb{M}^{n} \rightarrow
\op{Id}(A)$ is said to be proper if it maps only finitely many
monomials to proper ideals in $A$.     
%Let $\op{Id}(A)$ be the set of all ideals in the ring, $A$. Consider
%a coefficient ideal mapping from the set of all monomials to
%$\op{Id}(A)$ . If only finitely many monomials map to proper ideals
%in  $\op{Id}(A)$ then the mapping is called a proper coefficient
%ideal mapping. 
\end{definition}
%\begin{example}
%%Let $\mathfrak{a} = \langle 1, 3x, x^2, y\rangle$ be an ideal in $\mathbb{Z}[x,y]$. 
%Consider the mapping $\mathcal{I}: \mathbb{N}^2
%\longrightarrow \mathrm{Id}(\mathbb{Z} )$. Let $\mathcal{I} (1) = \{0\}, \mathcal{I} ({x_1}) = \langle 3 \rangle$
%and every other monomial be mapped to $\langle 1 \rangle$. We see that for any $x^\alpha \in \mathbb{N}^2$ such that $x_1 \mid x^\alpha$, 
%$\langle 3 \rangle \subseteq \mathcal{I}(x^\alpha)$.  Therefore, $\mathcal{I}$ is a proper coefficient ideal mapping. 
%\end{example}
%An example for order function is given in Appendix~\ref{appendix_label}.
\begin{example}
%Let $\mathfrak{a} = \langle 1, 3x, x^2, y\rangle$ be an ideal in $\mathbb{Z}[x,y]$. 
Consider the mapping $\mathcal{I}: \mathbb{N}^2
\longrightarrow \mathrm{Id}(\mathbb{Z} )$. Let $\mathcal{I} (1) = \{0\}, \mathcal{I} ({x_1}) = \langle 3 \rangle$
and every other monomial be mapped to $\langle 1 \rangle$. We see that for any $x^\alpha \in \mathbb{N}^2$ such that $x_1 \mid x^\alpha$, 
$\langle 3 \rangle \subseteq \mathcal{I}(x^\alpha)$.  Therefore, $\mathcal{I}$ is a proper order function. 
\end{example}
We define order ideal with respect to a proper order function $\mathcal{I}$. 
\begin{definition}
%Further examples for order function are given in Appendix~\ref{appendix_label}.
\label{Definition:OrderIdeal_Ring}  
For each $x^\alpha \in \mathbb{M}^n$ and a proper order function $\mathcal{I}$, fix $C_{x^\alpha}\subseteq A$, a set of coset representatives of $A/\mathcal{I}_{x^{\alpha}}$.
A set of terms $\mathcal{O}_{\mathcal{I}} \subseteq \mathbb{T}^{n}$ is said to be an order ideal w.r.t. $\mathcal{I}$ if for all $x^\alpha \in \mathbb{M}^n$, 
$cx^{\alpha} \in \mathcal{O}_{\mathcal{I}}$ if and only if 
    $c\in C_{x^\alpha}$.
\end{definition}
Note that for each monomial $x^\alpha$, one can choose any set of coset representatives of $A/\mathcal{I}_{x^{\alpha}}$ and with each choice we have a different order ideal. 
\begin{example} \label{exampleref}
  Consider the polynomial ring $\mathbb{Z}[x,y]$ and let the mapping $\mathcal{I}$ be such that $\mathcal{I}_1 = \{0\}$, $\mathcal{I}_x = \langle 4 \rangle$, $\mathcal{I}_y = \langle 3 \rangle$,
  $\mathcal{I}_{x^2} = \langle 2 \rangle$ and the rest of the monomials map to $\langle 1 \rangle$. $\mathcal{I}$ is clearly a proper order function. 
  Let the set of coset representatives be the following, $C_1 =  \mathbb{Z}, C_x = \{0,1,2,3\}, C_y = \{0,1,2\},C_{x^2} = \{0,1\}$ and for all the other monomials, $x^\alpha$, $C_{x^\alpha} = \{0\}$.
  Then, $\mathcal{O}_{\mathcal{I}} =$ $\{a_1,a_2x,a_3y,a_4x^{2}\ |$ $\ a_1 \in C_1,\ a_2 \in C_x,\ a_3 \in C_y,\  a_4 \in C_{x^2}\}$
 is an order ideal corresponding to $\mathcal{I}$.
\end{example}
\begin{example} \label{exampleref2}
 Now consider a polynomial ring \\$\Bbbk[u_1,u_2][x,y,z]$ and let $\mathcal{I}$ be an order function defined by 
 $\mathcal{I}_1 = \{0\}$, $\mathcal{I}_x = \{0\}$, $\mathcal{I}_y = \langle u_{1}^{2} \rangle$, $\mathcal{I}_z = \langle u_{2}^{2} - 3u_1 \rangle$, $\mathcal{I}_{x^{2}} = \langle 0 \rangle$, 
 $\mathcal{I}_{xy} = \langle u_{1}^{2}, u_{2}^{2}-1 \rangle$, $\mathcal{I}_{xz} = \langle u_{1}, u_2 \rangle$ and rest of the monomials mapping to $\langle 1 \rangle$. $\mathcal{I}$ is a proper order function. 
Let $C_1,C_x,C_y,C_z,C_{x^2}, C_{xy}, C_{xz}$ represent the nonzero set of coset representatives.
%  Then $\mathcal{O'}_{\mathcal{I}}=\{a_1,a_2x,a_3y,a_4z,a_5x^2,a_6xy,a_7xz\  |\ a_1 \in \Bbbk[u_1,u_2],\ a_2 \in \Bbbk[u_1,u_2],\ a_3 \in \Bbbk[u_1,u_2]/\langle u_{1}^{2} \rangle,
%  \ a_4 \in \Bbbk[u_1,u_2]/\langle u_{2}^{2} - 3u_1 \rangle,\ a_5 \in \Bbbk[u_1,u_2],\ a_6 \in \Bbbk[u_1,u_2]/\langle u_{1}^{2} , u_{2}^{2}-1 \rangle,\ 
%   a_7 \in \Bbbk[u_1,u_2]/\langle u_{1} , u_2  \rangle \}$ is an order ideal corresponding to $\mathcal{I}$.
%   
  Then $\mathcal{O'}_{\mathcal{I}}=\{a_1,a_2x,a_3y,a_4z,a_5x^2,$ $a_6xy,a_7xz\  |\ a_1 \in C_1,\ a_2 \in C_x,\ a_3 \in C_y,
 \ a_4 \in C_z,\ a_5 \in C_{x^2},\ a_6 \in C_{xy} ,\ 
  a_7 \in C_{xz} \}$ is an order ideal corresponding to $\mathcal{I}$.
\end{example}
% Note that the order ideal definition in the case of rings 
% (Definition~\ref{Definition:OrderIdeal_Ring}) is consistent with the
% order ideal definition in the case of fields (Definition~\ref{Definition:OrderIdeal_Field}).
In sequel, we write order ideal $\mathcal{O}_{\mathcal{I}}$
as $\mathcal{O}$ and its dependence on the order function is
implicitly assumed. 
It is important to note that unlike in the case of fields, the order ideal in the case of polynomial
rings over rings have both monic and nonmonic monomials.  
%Refer to Appendix~\ref{appendix_label} for more examples of order ideals.
% Therefore we
% introduce `monomial order ideal' $\Mon({\mathcal{O}})$ of an order ideal
% $\mathcal{O}$ defined as 
% ${\Mon(\mathcal{O}})=\{x^{\alpha}\ |\ cx^{\alpha} \in \mathcal{O}\}$

%Let $\mathcal{O}$ be an order ideal. The $\mathcal{O}$-monomial associated with $\mathcal{O}$ is %the set of all monomials in 
% $\mathcal{O}$. It is denoted as $\mathcal{O}_{m}$. Thus we have,
%  \begin{displaymath}
%  \mathcal{O}_{m}=\{x^{\alpha}\ |\ c.x^{\alpha} \in \mathcal{O}\ where\ c \in A/\mathcal{I}_{x^{\alpha}}\}.
%  \end{displaymath}
%\end{definition}
Given an order ideal $\mathcal{O}$ we introduce two types of borders: a
monomial border $\partial {\mathcal{O}}_{m}$ and a scalar border $\partial {\mathcal{O}}_{s}$.
\begin{definition}
Given an order ideal $\mathcal{O}$ the monomial border of
$\mathcal{O}$ is defined as  
\begin{displaymath}
 \partial\mathcal{O}_{m}=\{x_{1}\cdot\Mon(\mathcal{O}) \cup \ldots \cup
   x_{n}\cdot\Mon(\mathcal{O})\} \setminus \Mon(\mathcal{O}).  
 \end{displaymath}
\end{definition}
%In polynomial rings over rings there is another border associated
%with the order ideal that captures the nonmonic property of the order ideal. 
%\begin{definition}
%The scalar border of an order ideal $\mathcal{O}$ is defined as,
% \begin{displaymath}
% \partial\mathcal{O}_{s} = \{c_{1}x^{\alpha},\ldots,c_{s}x^{\alpha}\ |\ (\mathcal{I}_{x^{\alpha}},x^{\alpha}) \in 
% B,\ \mathcal{I}_{x^{\alpha}} = \langle c_{1},\ldots,c_{s} \rangle\}.
% \end{displaymath}
%\begin{displaymath}
%\partial \mathcal{O}_{s} = \underset{(\mathcal{I}_{x^{\alpha}},x^{\alpha}) \in
%  B}{\underset{x^{\alpha} \in \mathbb{T}^{n}}{\bigcup}} \{ c_{1}x^{\alpha},\ldots,c_{s}x^{\alpha} \}
%\end{displaymath}  
%%such that the ideal membership problem can be verified for $\mathcal{I}_{x^{\alpha}}$ through its generators.
%\end{definition}
%Scalar Border of an Order Ideal
\begin{definition}
Let $\mathcal{O}$ be an order ideal with respect to a proper
order function $\mathcal{I}$. For each $x^{\alpha}$ such 
that $\mathcal{I}_{x^{\alpha}} \neq \langle 1 \rangle$ define
$ \partial \mathcal{O}_{x^{\alpha}} = \{ c_{1}x^{\alpha},\ldots,c_{s}x^{\alpha} \}$,
where $\mathcal{I}_{x^{\alpha}} = \langle c_{1},\ldots, c_{s}\rangle$ for some
$c_{1},\ldots,c_{s} \in A$.
The scalar border of an order ideal is defined as 
\begin{displaymath}
\partial \mathcal{O}_{s} = \underset{\mathcal{I}_{x^{\alpha}} \neq \langle 1 \rangle}{\underset{x^{\alpha} \in \mathbb{M}^{n}}{\bigcup}} \partial \mathcal{O}_{x^{\alpha}}.
\end{displaymath}  
%such that the ideal membership problem can be verified for $\mathcal{I}_{x^{\alpha}}$ through its generators.
\end{definition}

\begin{definition}\label{borderofideal}
The border of the order ideal $\mathcal{O}$, denoted as
$\partial\mathcal{O}$, is defined as $\partial\mathcal{O} = \partial
\mathcal{O}_{m} \cup \partial \mathcal{O}_{s}$. 
\end{definition}
\begin{example}
 Consider Example~\ref{exampleref}.
 The set of monic border terms that form the monomial border is
 $\partial\mathcal{O}_{m}=\{xy,y^{2},$ $x^{3},x^{2}y\}$ and  
 the scalar border is
 $\partial\mathcal{O}_{s}=\{4x,3y,2x^{2}\}$. Hence, the border of the order
 ideal  
 %$\mathcal{O}=\{a_1,a_2x,a_3y,a_4x^{2}\ |\ a_1 \in \mathbb{Z},\ a_2 \in
 %\mathbb{Z}_{4},\ a_3 \in \mathbb{Z}_{3},\ a_4 \in \mathbb{Z}_{2}\}$ 
 is
 $\partial \mathcal{O} = \{ xy,y^{2},x^{3},x^{2}y,4x,3y,2x^{2} \}$.
\end{example}

\begin{example}
 Consider Example~\ref{exampleref2}.
 The monomials that form the monomial border is
 $\partial\mathcal{O'}_{m}=\{x^3,x^2y,x^2z,xy^2,$ $xyz,xz^2,y^2,yz,z^2\}$ and  
 the scalar border is the set
 $\partial\mathcal{O'}_{s}=\{u_{1}^{2}y,(u_{2}^{2}-3u_1)z,u_{1}^{2}xy,(u_{2}^{2}-1)xy,u_1xz,u_2xz\}$. Hence border of the order
 ideal  
 %$\mathcal{O'} = \{a_1,a_2x,a_3y,a_4z,a_5x^2,a_6xy,a_7xz \  |\ a_1 \in \Bbbk[u_1,u_2],\ a_2 \in \Bbbk[u_1,u_2],\ a_3 \in \Bbbk[u_1,u_2]/\langle u_{1}^{2} \rangle,
 %\ a_4 \in \Bbbk[u_1,u_2]/\langle u_{2}^{2} - 3u_1 \rangle,\ a_5 \in \Bbbk[u_1,u_2],\ a_6 \in \Bbbk[u_1,u_2]/\langle u_{1}^{2} , u_{2}^{2}-1 \rangle,\ 
  %a_7 \in \Bbbk[u_1,u_2]/\langle u_{1} , u_2  \rangle \}$ 
  is
 $\partial \mathcal{O'} = \{ x^3,x^2y,x^2z,xy^2,xyz,xz^2,y^2,$ $yz,z^2,u_{1}^{2}y,(u_{2}^{2}-3u_1)z$, $u_{1}^{2}xy,(u_{2}^{2}-1)xy,u_1xz,u_2xz \}$.
\end{example}
%\begin{example}
% Consider Example~\ref{exampleref}.
% The set of monic border terms that form the monomial border is
% $\partial\mathcal{O}_{m}=\{xy,y^{2},$ $x^{3},x^{2}y\}$ and  
% the scalar border is
% $\partial\mathcal{O}_{s}=\{4x,3y,2x^{2}\}$. Hence, the border of the order
% ideal  
% %$\mathcal{O}=\{a_1,a_2x,a_3y,a_4x^{2}\ |\ a_1 \in \mathbb{Z},\ a_2 \in
% %\mathbb{Z}_{4},\ a_3 \in \mathbb{Z}_{3},\ a_4 \in \mathbb{Z}_{2}\}$ 
% is
% $\partial \mathcal{O} = \{ xy,y^{2},x^{3},x^{2}y,4x,3y,2x^{2} \}$.
%\end{example}
%
%\begin{example}
% Consider Example~\ref{exampleref2}.
% The monomials that form the monomial border is
% $\partial\mathcal{O'}_{m}=\{x^3,x^2y,x^2z,xy^2,$ $xyz,xz^2,y^2,yz,z^2\}$ and  
% the scalar border is the set
% $\partial\mathcal{O'}_{s}=\{u_{1}^{2}y,(u_{2}^{2}-3u_1)z,u_{1}^{2}xy,(u_{2}^{2}-1)xy,u_1xz,u_2xz\}$. Hence border of the order
% ideal  
% %$\mathcal{O'} = \{a_1,a_2x,a_3y,a_4z,a_5x^2,a_6xy,a_7xz \  |\ a_1 \in \Bbbk[u_1,u_2],\ a_2 \in \Bbbk[u_1,u_2],\ a_3 \in \Bbbk[u_1,u_2]/\langle u_{1}^{2} \rangle,
% %\ a_4 \in \Bbbk[u_1,u_2]/\langle u_{2}^{2} - 3u_1 \rangle,\ a_5 \in \Bbbk[u_1,u_2],\ a_6 \in \Bbbk[u_1,u_2]/\langle u_{1}^{2} , u_{2}^{2}-1 \rangle,\ 
%  %a_7 \in \Bbbk[u_1,u_2]/\langle u_{1} , u_2  \rangle \}$ 
%  is
% $\partial \mathcal{O'} = \{ x^3,x^2y,x^2z,xy^2,xyz,xz^2,y^2,$ $yz,z^2,u_{1}^{2}y,(u_{2}^{2}-3u_1)z$, $u_{1}^{2}xy,(u_{2}^{2}-1)xy,u_1xz,u_2xz \}$.
%\end{example}
We define $\partial^{0}\mathcal{O} = \mathcal{O}$ and $0^{th}$ border
closure as
%\begin{displaymath}
 $   \overline{\partial^{0}\mathcal{O}} = \{
ax^{\alpha}\ |\ cx^{\alpha} \in \mathcal{O}, 
    \text{ for some }  c \in C_{x^{\alpha}}, c \neq 0 \text{ and } a \in c + \mathcal{I}_{x^{\alpha}}\}$.
%\end{displaymath}
Note that in the case of fields these quantities are defined
as $\partial^{0}\mathcal{O} =
\overline{\partial^{0}\mathcal{O}} = \mathcal{O}$ ~\citep{KehreinKreuzer:2006:ComputingBorderBases}.

The definitions of first and higher order  border closures are given below. 
\begin{definition}%[$1^{st}$ border closure]
 \label{Definition:FirstBorderClosure} 
The first border closure $\overline{\partial\mathcal{O}}$ of an order
ideal $\mathcal{O}$ is defined as 
\begin{align*}
  \overline{\partial\mathcal{O}} = &\{ax^{\alpha}\ |\ \exists c \in A
\ \text{such that} \ cx^{\alpha} \in \mathcal{O} \cup
\partial\mathcal{O}_{s}, \\ &a \in A \text{ or
} x^{\alpha} \in \partial\mathcal{O}_{m}\}.  
\end{align*}
\end{definition}

\begin{proposition}
The first border closure, $\overline{\partial\mathcal{O}}$,  of an
order ideal, $\mathcal{O}$ is an order ideal. 
\end{proposition}
\begin{proof}
%By definition,
%\begin{displaymath}
%\overline{\partial\mathcal{O}} = \{ax^{\alpha}\ |\ a \in
%A,\ cx^{\alpha} \in \mathcal{O} \cup \partial\mathcal{O}_{s} \text{ or
%} x^{\alpha} \in \partial\mathcal{O}_{m}\}.  
%\end{displaymath}
We fix $\mathcal{I}_{x^{\alpha}} = \{0\}$ for all $x^{\alpha} \in
\overline{\partial\mathcal{O}}$. 
%We prove that
%$\overline{\partial\mathcal{O}}$ is an order ideal.
By Definition~\ref{Definition:FirstBorderClosure} three cases arises:
$x^{\alpha} \in \partial\mathcal{O}_{m}$ or  there exists $c \in A$
such that $cx^{\alpha} \in \mathcal{O}$ or $cx^{\alpha} \in \partial\mathcal{O}_{s}$.
Let $x^{\alpha} \in \partial\mathcal{O}_{m}$. Suppose $x^{\beta} |
x^{\alpha}$ for some $x^{\beta} \in \mathbb{M}^{n}$, 
then clearly, $x^{\beta} \in \Mon(\mathcal{O}) \cup
\partial\mathcal{O}_{m}$. If $x^{\beta} \in \Mon(\mathcal{O})$, then
there exists some $d\in C_{x^{\beta}}$ such that $dx^{\beta}
\in \mathcal{O}$. Therefore, in either case $x^{\beta} \in
\overline{\partial\mathcal{O}}$. In the second case, let $cx^{\alpha} \in
\mathcal{O}$. Suppose $x^{\beta} | x^{\alpha}$ for some $x^{\beta} \in
\mathbb{M}^{n}$. By the closure property of $\mathcal{O}$, $dx^{\beta} \in
\mathcal{O}$ for some $d \in C_{x^{\beta}}$. Therefore, $x^{\beta}
\in \overline{\partial\mathcal{O}}$. In the third case, let
$cx^{\alpha} \in \partial\mathcal{O}_{s}$. Suppose $x^{\beta} | x^{\alpha}$
for some $x^{\beta} \in \mathbb{M}^{n}$. This implies that $x^{\beta} \in \mathrm{Mon}(\mathcal{O})$. Thus, $x^{\beta} \in
\overline{\partial\mathcal{O}}$.  
\end{proof}

The monomial part of the first border closure defined as the set of monomials in $\overline{\partial\mathcal{O}}$, 
is a finite set and it is represented as $\Mon(\overline{\partial\mathcal{O}})$. It is interesting to see that since $\mathcal{I}_{x^{\alpha}} = \{0\}$ for all 
$x^{\alpha} \in \mathrm{Mon}(\overline{\partial\mathcal{O}})$, the scalar border for $k \geq 2$ is an empty set and one needs to consider
only the monomial border.
\begin{definition}\label{kthborder}
 The $k^{th}$ border of an order ideal $\mathcal{O}$ for $k \gneq 1$
 is defined as
 \begin{displaymath}
 \partial^{k}\mathcal{O} =
 \{x_{1}\cdot\Mon(\overline{\partial^{k-1}\mathcal{O}}) \cup \ldots \cup
 x_{n}\cdot\Mon(\overline{\partial^{k-1}\mathcal{O}})\} \setminus
 \Mon(\overline{\partial^{k-1}\mathcal{O}}), 
 \end{displaymath}
 where $\Mon(\overline{\partial^{k-1}\mathcal{O}})$ is the monomial part of the $(k-1)^{th}$ border closure.
\end{definition}

\begin{definition}
For $k \geq 2$, the $k^{th}$ border closure of an order ideal is
defined as
\begin{displaymath}
  \overline{\partial^{k}\mathcal{O}} = \{ax^{\alpha} |\ a \in A, x^{\alpha} \in \partial^{k}\mathcal{O} \cup \overline{\partial^{k-1}\mathcal{O}_{m}}\}. 
\end{displaymath}
\end{definition}
\begin{example}\label{borderexample}
Consider Example~\ref{exampleref}.
The set of monic border terms that form the monomial border is
$\partial\mathcal{O}_{m}=\{xy,y^{2},$ $x^{3},x^{2}y\}$ and the scalar
border is $\partial\mathcal{O}_{s}=\{4x,3y,2x^{2}\}$. 
The second border of the order ideal, $\mathcal{O}$, is the set, 
$\partial^2\mathcal{O}= \{xy^2,y^3,$ $x^4,x^3y,x^2y^2\}$.
\end{example}

\begin{example}\label{borderexample2}
Consider Example~\ref{exampleref2}. The second border of the order ideal, $\mathcal{O'}$ is the set, $\partial^{2}\mathcal{O'} = \{ y^3, y^2z,yz^2, z^3, x^4, $ $x^3y, x^3z, x^2y^2, x^2yz, 
x^2z^2, xy^3, xyz^2, xy^2z,xz^3\}$.
\end{example}
%\begin{example}\label{borderexample}
%Consider Example~\ref{exampleref}.
%The set of monic border terms that form the monomial border is
%$\partial\mathcal{O}_{m}=\{xy,y^{2},$ $x^{3},x^{2}y\}$ and the scalar
%border is $\partial\mathcal{O}_{s}=\{4x,3y,2x^{2}\}$. 
%The second border of the order ideal, $\mathcal{O}$, is the set, 
%$\partial^2\mathcal{O}= \{xy^2,y^3,$ $x^4,x^3y,x^2y^2\}$.
%\end{example}
%
%\begin{example}\label{borderexample2}
%Consider Example~\ref{exampleref2}. The second border of the order ideal, $\mathcal{O'}$ is the set, $\partial^{2}\mathcal{O'} = \{ y^3, y^2z,yz^2, z^3, x^4, $ $x^3y, x^3z, x^2y^2, x^2yz, 
%x^2z^2, xy^3, xyz^2, xy^2z,xz^3\}$.
%\end{example}

\begin{remark}
 The $k^{th}$ border closure is an infinite set of terms for $k \geq 0$. Further, for $k \geq 1$, 
 $\overline{\partial^{k}\mathcal{O}}$ is closed under division and hence the set of monomials corresponding to it, 
 $\Mon(\overline{\partial^{k}\mathcal{O}})$, mimics the case of fields.
\end{remark}
 %Appendix~\ref{appendix_label} provides examples and figures that explain the borders and  border closures of an order ideal.
 The following example  explains the borders and  border closures of an order ideal.
 \begin{example}
Let the order function $\mathcal{I}: \mathbb{M}^{2} \rightarrow \op{Id}(\mathbb{Z})$ be defined as follows: $\mathcal{I}_1 = 0$,  $\mathcal{I}_x = 0$, $\mathcal{I}_y = \langle 4 \rangle$, $\mathcal{I}_{xy} = \langle 2 \rangle$ and for other monomials $\mathcal{I}$ is mapped to $\langle 1 \rangle$. The order ideal corresponding to $\mathcal{I}$, $\mathcal{O}_\mathcal{I} = \{a, bx, cy, dxy : a,b \in \mathbb{Z}, c \in \{0,1,2,3\}, d \in \{0,1\} \}$. The $0^{th}$ border closure is given by $\{a,bx,cy,dxy : a, b \in \mathbb{Z}\setminus  \{0\}, c \in \mathbb{Z}\setminus\langle 4 \rangle, d \in \mathbb{Z} \setminus \langle 2 \rangle \}$. Figure~\ref{ZeroOrderclosure} shows a few terms from the $0^{th}$ border closure. 
\begin{figure}[h!]
%\centering
\includegraphics[width=60mm]{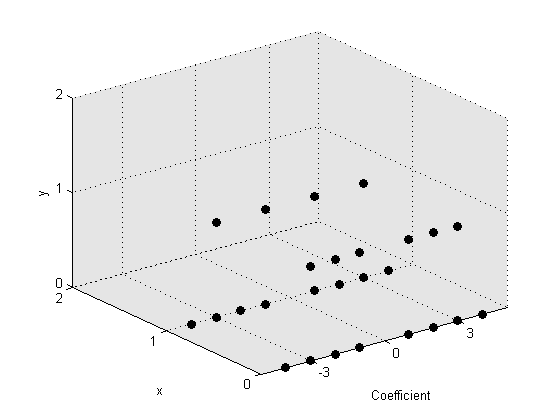}
\caption{$0^{th}$ border closure\label{overflow}} \label{ZeroOrderclosure}
\end{figure}\\
The first border is $\langle x^2, 2xy, y^2, x^2y, xy^2, 4y \rangle$. The first border closure is given by $\langle ax^2, bxy, cy^2, dx^2y, exy^2, fy : a,c,d,e \in \mathbb{Z}, b \in \langle 2 \rangle, f \in \langle 4  \rangle$. Figure \ref{firstborder} is an illustration of the same. 
\begin{figure}[h!]
%\centering
\includegraphics[width=60mm]{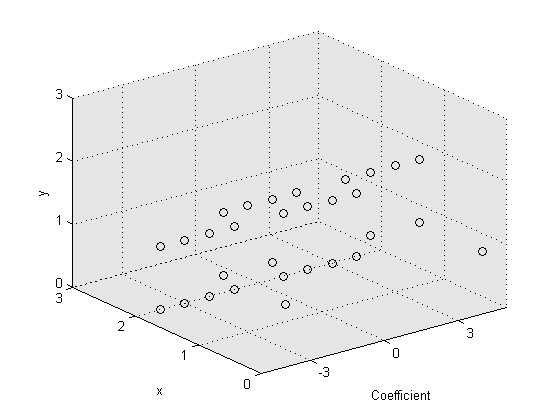}
\caption{First border closure\label{overflow}} \label{firstborder}
\end{figure}\\
For $k \geq 2$ the $k^{th}$ borders are exactly as that of fields. Figure \ref{otherborders} shows the borders  for $k=2,3$. Note that the figure depicts the borders and not the border closures. 
\begin{figure}[h!]
%\centering
\includegraphics[width=60mm]{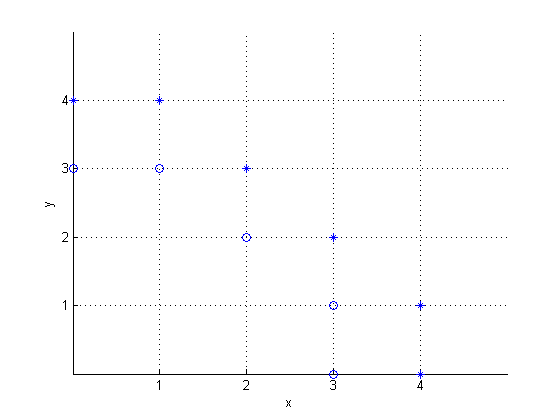}
\caption{$2^{nd}, 3^{rd}$ border closure\label{overflow}} \label{otherborders}
\end{figure}
\end{example}
We give below certain properties of order ideals, their borders and border closures. These properties are analogous to the case of polynomial rings over fields. 
\begin{proposition}
  \label{BorderProp}
 Let $\mathcal{O}$ be an order ideal and $\partial\mathcal{O}_{m}$ be
 its monomial border. Then 
 \begin{enumerate}
  \item For $k \geq 1$, the $k^{th}$ monomial border closure of $\mathcal{O}$, $\Mon(\overline{\partial^{k}\mathcal{O}})$ is the following disjoint
  union, $\Mon(\mathcal{O}) \cup \partial\mathcal{O}_{m} \cup (\cup_{i=2}^{k} \partial^{k}\mathcal{O})$.
  \item For $k \geq 1$, $\partial^{k}\mathcal{O}_{m}$ = $\mathbb{M}_{k}^{n}\cdot \Mon(\mathcal{O}) \setminus \mathbb{M}_{<k}^{n}.\Mon(\mathcal{O})$,
  where $\partial^{k}\mathcal{O}_m=\partial^{k}\mathcal{O}$ for $k \geq 2$.
  \item A monomial, $x^{\alpha} \in \mathbb{M}^{n}$ is divisible by $x^{\beta} \in \partial\mathcal{O}_{m}$ if and only if 
  $x^{\alpha} \in \mathbb{M}^{n} \setminus \Mon(\mathcal{O})$.
 \end{enumerate}
\end{proposition}
\begin{proof}
%\begin{enumerate}
 %\item 
(1) The proof is by induction on $k$. For $k = 1$, clearly the monomials in the 
first border closure are elements of the set 
$\mathcal{O}_{m} \cup \partial\mathcal{O}_{m}$. From the definition of monomial border of $\mathcal{O}$ we have that
$\mathcal{O}_{m}$ and $\partial\mathcal{O}_{m}$ are disjoint. Suppose
that the claim is true for the $k^{th}$ monomial border  
closure. For $k+1$, $\overline{\partial^{k+1}\mathcal{O}_{m}}= \overline{\partial^{k}\mathcal{O}_{m}} \cup \partial^{k+1}\mathcal{O}$.
It is easy to verify that the sets $\overline{\partial^{k}\mathcal{O}_{m}}$ and $\partial^{k+1}\mathcal{O}$ are disjoint.\\
%\item
(2) The claim follows from the observation that $\partial^{k}\mathcal{O}$ = $\overline{\partial^{k}\mathcal{O}_{m}} \setminus 
\overline{\partial^{k-1}\mathcal{O}_{m}}$. \\
%\item
(3) We have $x^{\beta} \in \partial\mathcal{O}_{m}$. This implies that there exists $x^{\gamma} \in \mathcal{O}_{m}$ and an indeterminate
$x_{i_{0}}$ such that $x^{\beta} = x_{i_{0}}x^{\gamma}$. 
We have $x_{i_{0}}x^{\gamma} | x^{\alpha}$. If $x^{\alpha} \in \mathcal{O}_{m}$ then $x_{i_{0}}x^{\gamma} \in \mathcal{O}_{m}$ which is
a contradiction. Now consider a monomial $x^{\alpha} \in \mathbb{T}^{n} \setminus \mathcal{O}$. Then, $x^{\alpha} \in \partial\mathcal{O}_{m}$
or $x^{\alpha} \in \partial^{k}\mathcal{O}$ for some $k \geq 2$. If 
$x^{\alpha} \in \partial^{k}\mathcal{O}$ then it implies that there exists a monomial $x^{\gamma}$ of degree $k-1$ and a $x^{\beta} \in 
\partial\mathcal{O}_{m}$ such that $x^{\alpha} =
x^{\gamma}x^{\beta}$. The claim follows. 
%\end{enumerate}
\end{proof}

Now we introduce some concepts that are essential for the division
algorithm.
%Definition: Index of a Term
\begin{definition}
The index of a term $cx^{\alpha}$ w.r.t. an order ideal, $\mathcal{O}$ is
defined as 
\begin{displaymath}
\operatorname{ind_{\mathcal{O}}}(cx^{\alpha}) =
\min \{k \in \mathbb{N}\ |\ cx^{\alpha} \in
\overline{\partial^{k}\mathcal{O}}\}.
\end{displaymath}
\end{definition}

%Definition: Index of a Polynomial
\begin{definition}\label{Index definition}
Let $f \in A[x_{1},\ldots,x_{n}]$ 
be any  nonzero polynomial with support, $\op{supp}(f)$, then the index of $f$ w.r.t. an order ideal, $\mathcal{O}$ is defined as 
\begin{displaymath}
\op{ind_{\mathcal{O}}}(f) = \underset{\alpha \in \op{supp}(f)}{\op{max}}
\op{ind_{\mathcal{O}}}(c_{\alpha} x^{\alpha}).
\end{displaymath}   
\end{definition}
\begin{example}
\label{index_example}
 Consider Example~\ref{exampleref}. 
 %The order ideal 
% $\mathcal{O}=\{a_1,a_2x,a_3y,a_4x^{2}\ |\ a_1 \in \mathbb{Z},\ a_2 \in 
 %\mathbb{Z}_{4},\ a_3 \in \mathbb{Z}_{3},\ a_4 \in \mathbb{Z}_{2}\}$.
 The set of monic border terms that form the monomial border is $\partial\mathcal{O}_{m}=\{xy,y^{2},$ $x^{3},x^{2}y\}$ and 
 the scalar border is $\partial\mathcal{O}_{s}=\{4x,3y,2x^{2}\}$. Then
 $\operatorname{ind}_{\mathcal{O}}(3x)=0$, 
$\operatorname{ind}_{\mathcal{O}}(xy)=1$ and
 $\operatorname{ind}_{\mathcal{O}}(xy^2+8x+7y)=2$. 
\end{example}

\begin{example}
\label{index_example2}
 Consider Example~\ref{exampleref2}. 
 %The order ideal 
 %$\mathcal{O'} = \{a_1,\ a_2x,\ a_3y,\ a_4z,$ $a_5x^2,\ a_6xy,\ a_7xz\  |\ a_1 \in \Bbbk[u_1,u_2],\ a_2 \in \Bbbk[u_1,u_2],\ a_3 \in \Bbbk[u_1,u_2]/\langle u_{1}^{2} \rangle,\ 
  %a_4 \in \Bbbk[u_1,u_2]/\langle u_{2}^{2} - 3u_1 \rangle,\ a_5 \in \Bbbk[u_1,u_2],\ a_6 \in \Bbbk[u_1,u_2]/\langle u_{1}^{2} , u_{2}^{2}-1 \rangle,\ 
  %a_7 \in \Bbbk[u_1,u_2]/\langle u_{1} , u_2  \rangle \}$.  
  The monomials that form the monomial border is $\partial\mathcal{O'}_{m}=\{x^3,x^2y,x^2z,xy^2,$ $xyz,xz^2,y^2,yz,z^2\}$ and  
 the scalar border is the set $\partial\mathcal{O'}_{s}=\{u_{1}^{2}y,(u_{2}^{2}-3u_1)z,u_{1}^{2}xy,(u_{2}^{2}-1)xy,u_1xz,u_2xz\}$. Then \\
 $\operatorname{ind}_{\mathcal{O'}}((3u_2+u_2)x)=0$, 
$\operatorname{ind}_{\mathcal{O'}}(u_1x^2y)= \operatorname{ind}_{\mathcal{O'}}((u_{1}^{2}+u_2)xz) = 1$ and
 $\operatorname{ind}_{\mathcal{O'}}(xy^3+8x)=2$. 
\end{example}
For any polynomial, the terms of highest index are grouped together to
form a border form analogous 
to the leading term in Gr\"obner bases theory. We define this below. 
\begin{definition}\label{border form definition}
 Let $f \in A[x_1, \ldots, x_n]$ be a nonzero polynomial such that the $\operatorname{ind_{\mathcal{O}}}(f)$ = $i_{0}$.
 The border form of $f$ w.r.t. $\mathcal{O}$ is defined as 
\begin{displaymath}
   \op{BF}_{\mathcal{O}}(f) = \underset{\operatorname{ind_{\mathcal{O}}}(c_{\alpha}x^{\alpha}) = i_{0}}{ \underset{\alpha \in \op{supp}(f), c_\alpha \in A}{\sum}} 
   c_{\alpha}x^{\alpha} \enspace,
\end{displaymath}
a polynomial in $A[x_{1},\ldots,x_{n}]$.
\end{definition}
 Note that unlike leading term of a polynomial in Gr\"obner bases theory that is always a monomial, border form can be a polynomial. The concept of leading term  ideal has an analogous form in border bases theory called the 
border form ideal.
%Definition:Border form Ideal
\begin{definition}
%\label{border form ideal definition}
The border form ideal of an ideal $\mathfrak{a}$ in $A[x_1, \ldots, x_n]$ w.r.t. an order ideal $\mathcal{O}$ is defined as
\begin{displaymath}
  \op{BF}_{\mathcal{O}}(\mathfrak{a}) = \langle
  \op{BF}_{\mathcal{O}}(f)\ |\ f \in \mathfrak{a} \rangle.
\end{displaymath}
\end{definition}
\begin{example}\label{border_form_example}
Consider Example ~\ref{exampleref}. Let $f=xy^2+2x^2y^2+xy+3x+2$. Then the index of $f$ w.r.t. the order ideal $\mathcal{O}$ is 
equal to 2. The border form of $f$ is the polynomial, $\operatorname{BF}_{\mathcal{O}}(f)=xy^2+2x^2y^2$.
\end{example}

\begin{example}\label{border_form_example2}
Consider Example ~\ref{exampleref2}. Let $f=2xy^2+(u_1-u_2)xz+3x+2$. Then the index of $f$ w.r.t. the order ideal $\mathcal{O'}$ is 
equal to 2. The border form of $f$ is the polynomial, $\operatorname{BF}_{\mathcal{O'}}(f)=2xy^2+(u_1-u_2)xz$.
\end{example}
%\begin{example}\label{border_form_example}
%Consider Example ~\ref{exampleref}. Let $f=xy^2+2x^2y^2+xy+3x+2$. Then the index of $f$ w.r.t. the order ideal $\mathcal{O}$ is 
%equal to 2. The border form of $f$ is the polynomial, $\operatorname{BF}_{\mathcal{O}}(f)=xy^2+2x^2y^2$.
%\end{example}
%
%\begin{example}\label{border_form_example2}
%Consider Example ~\ref{exampleref2}. Let $f=2xy^2+(u_1-u_2)xz+3x+2$. Then the index of $f$ w.r.t. the order ideal $\mathcal{O'}$ is 
%equal to 2. The border form of $f$ is the polynomial, $\operatorname{BF}_{\mathcal{O'}}(f)=2xy^2+(u_1-u_2)xz$.
%\end{example}

We now give the definition of border prebasis for an order ideal,  $\mathcal{O}$.
%Border Prebasis
\begin{definition}
Let $\mathcal{O}$ be an order ideal, and
$\partial\mathcal{O}=\{c_{1}x^{\alpha_{1}},\ldots,c_{s}x^{\alpha_{s}}\}$
be its border. Let $C_{x^{\alpha_{i}}}$ be the set of coset 
representatives of $A/\mathcal{I}_{x^{\alpha_{i}}}$.
A finite set of polynomials $G$ = $\{g_{1},\ldots,g_{s}\} \subseteq
A[x_{1},\ldots,x_{n}]$ is said to be an $\mathcal{O}$-border prebasis
if $g_{i}=c_{i}x^{\alpha_{i}}- h_{i}$,  where
$h_{i} \in A[x_{1},\ldots,x_{n}]$ satisfying
$\mathrm{Ter}(h_{i}) \subseteq \mathcal{O} \setminus \{ax^{\alpha_{i}}\ |\ a
\in C_{x^{\alpha_{i}}}\}$, $i = 1,\ldots,s$.  
\end{definition}
\begin{example}\label{border_prebases_example}
We consider Example~\ref{exampleref}. 
%We have the order ideal $\mathcal{O}=\{a_1,a_2x,a_3y,a_4x^{2}$ $|\ a_1 \in \mathbb{Z},\ a_2 \in \mathbb{Z}_{4},\ a_3 \in \mathbb{Z}_{3},\  a_4 \in \mathbb{Z}_{2}\}$ and the $\mathcal{O}$-border
%$\partial\mathcal{O}=\{xy,y^{2},x^{3},x^{2}y,4x,3y,2x^{2}\}$. 
The set $G=\{g_1,\ldots,g_7\}$, where $g_1=xy-x$, $g_2=y^2-y$, $g_3=x^3-2y$, $g_4=x^2y-x^2+10$,  $g_5=4x-2y$, $g_6=3y-3x$ and $g_7=2x^2-x+5$ is an 
$\mathcal{O}$-border prebasis but it is not acyclic. 
Let $G'=\{g'_1,\ldots,g'_7\}$ 
where $g'_1=xy-x$, $g'_2=y^2-y$, $g'_3=x^3-x^2+6$, $g'_4=x^2y-y+5$, $g'_5=4x-7$, $g'_6=3y-x$ and $g'_7=2x^2-2y-3x$ which is also a $\mathcal{O}$-border prebasis but it is acyclic since the permutation of $G'$, $\{g'_1,g'_2,g'_3,g'_4,g'_7,g'_6,g'_5\}$ satisfies  
the acyclicity condition.
\end{example}

\begin{example}\label{border_prebases_example2}
Consider Example~\ref{exampleref2}. 
%We have the order ideal $\mathcal{O'} = \{a_1,\ a_2x,\ a_3y,\ a_4z,$ $a_5x^2,\ a_6xy,\ a_7xz \  |\ 
%a_1 \in \Bbbk[u_1,u_2],\ a_2 \in \Bbbk[u_1,u_2],\ a_3 \in \Bbbk[u_1,u_2]/\langle u_{1}^{2} \rangle,
 %\ a_4 \in \Bbbk[u_1,u_2]/\langle u_{2}^{2} - 3u_1 \rangle,\ a_5 \in \Bbbk[u_1,u_2],\ a_6 \in \Bbbk[u_1,u_2]/\langle u_{1}^{2} , u_{2}^{2}-1 \rangle,\ 
 % a_7 \in \Bbbk[u_1,u_2]/\langle u_{1} , u_2  \rangle \}$ and the $\mathcal{O'}$-border, 
 %$\partial \mathcal{O'} = \{ x^3,x^2y,x^2z,xy^2,xyz,xz^2,y^2,yz,z^2,u_{1}^{2}y,(u_{2}^{2}-3u_1)z,u_{1}^{2}xy,(u_{2}^{2}-1)xy,u_1xz,u_2xz \}$. 
The set  $G =\\ \{g_1,\ldots,g_{15}\}$, where $g_1=x^3-3$, $g_2=x^2y-3u_1y$, $g_3=x^2z-2z$, $g_4=xy^2-x+10$,  $g_5=xyz-11xy$, $g_6=xz^2-u_{2}u_{1}^{2}x^2$, $g_7=y^2-x+u_1u_2$, $g_8=yz-3y+2$, 
$g_9=z^2+5xz+11u_1x$, $g_{10}=u_{1}^{2}y+u_2x+3$, $g_{11}=(u_{2}^{2}-3u_1)z-u_{2}^{2}y$, $g_{12}=u_{1}^{2}xy+3u_1x-2z$, $g_{13}=(u_{2}^{2}-1)xy+2x^2$, $g_{14}=u_1xz+3u_1x^2$, 
$g_{15}=u_2xz+2u_{1}xy+4x^2-4z-10u_1y+14$ 
is an acyclic $\mathcal{O}$-border prebasis since the following permutation of $G$, $\{g_1,g_2,g_3,g_4,g_5,g_6,g_7,g_8,g_9,g_{14},g_{15},g_{13},g_{12},g_{11},g_{10}\}$ satisfies  
the acyclicity condition.
\end{example}
%Examples for border form ideal and $\mathcal{O}$-border prebasis are given in Appendix~\ref{appendix_label}.
Note that unlike in fields, for a monomial in the border of $\mathcal{O}$, we can have more than one polynomial in the $\mathcal{O}$-border prebasis but only one polynomial corresponding to a term in the border.
With the definition of $\mathcal{O}$-border prebasis, we now give a procedure for division of any polynomial in $A[x_1,\ldots,x_n]$ with the $\mathcal{O}$-border prebasis.  
%Border Division Algorithm
\begin{algo}\label{ProcedureDiv}
Let $\mathcal{O}$ be an order ideal. Let $\Mon(\mathcal{O})=\{x^{\alpha_{1}},\ldots,x^{\alpha_{t}}\}$ 
be its monomial part. Let $\partial\mathcal{O}_{m}=\{x^{\beta_{1}},\ldots,x^{\beta_{s'}}\}$ and 
$\partial\mathcal{O}_{s}=\{c_{s'+1}x^{\beta_{s'+1}},\ldots,c_{s}x^{\beta_{s}}\}$ be its monomial border and scalar border respectively.
Let $G=\{g_{1},\ldots,g_{s}\} \subseteq A[x_{1},\ldots,x_{n}]$ be an $\mathcal{O}$-border prebasis.
For $f \in A[x_1,\ldots,x_n]$ we perform the following steps.

\begin{enumerate}
 \item Initialize $f_{1}=\ldots=f_{s}=0$, $a_{1}=\ldots=a_{t}=0$ and
       $h=f$. 
 \item If $h=0$ return ($f_{1},\ldots,f_{s},a_{1},\ldots,a_{t}$). 
 \item If $\op{ind}_{\mathcal{O}}(h)$ = $0$ then find
       $b_{1},\ldots,b_{t} \in A$ such that
       $h=b_{1}x^{\alpha_{1}}+\ldots+b_{t}x^{\alpha_{t}}$. 
 Set $a_{i}=b_{i}$ for each $1 \leq i \leq t$. Return ($f_{1},\dots,f_{s},a_{1},\ldots,a_{t}$).
\item If $\op{ind}_{\mathcal{O}}(h)$ = $1$ and $h$ contains a term
  $dx^{\beta}$ such that $x^{\beta} \in \partial\mathcal{O}_{m}$ then  
goto Step 5. Else, let $h$ =
$d_{1}x^{\gamma_1}+\ldots+d_{j}x^{\gamma_j}$ such that
$\op{ind}_{\mathcal{O}}(h) = \op{ind}_{\mathcal{O}}(d_{1}x^{\gamma_1})$
and $\op{ind}_{\mathcal{O}}(d_{1}x^{\gamma_1}) \geq \ldots \geq  
\op{ind}_{\mathcal{O}}(d_{j}x^{\gamma_j})$. Find
$b_{\mu+1},\ldots,b_{s} \in A$ such that
$d_{1}x^{\gamma_1}=b_{s'+1}(c_{s'+1}x^{\beta_{s'+1}})+\ldots+b_{s}(c_{s}x^{\beta_{s}})$. Subtract
$b_{s'+1}g_{s'+1}+\ldots+b_{s}g_{s}$ from $h$,  add
$b_{i}\ to\ f_{i}$ for $s'+1 \leq i \leq s$ and return to Step 2. 
\item Else, let $h$ = $d_{1}x^{\gamma_1}+ \ldots +d_{j}x^{\gamma_j}$ such that 
  $\op{ind}_{\mathcal{O}}(h) = \op{ind}_{\mathcal{O}}(x^{\gamma_1})$
  and $\op{ind}_{\mathcal{O}}(d_{1}x^{\gamma_1}) \geq \ldots \geq  
  \op{ind}_{\mathcal{O}}(d_{j}x^{\gamma_j})$. Determine $x^{\beta_{i}} \in
  \partial\mathcal{O}_{m}$ with the smallest $i$ 
  such that $x^{\gamma_i}=x^{\mu}x^{\beta_i}$ and $\op{deg}(x^{\mu})
  =\op{ind}_{\mathcal{O}}(h)-1$.  
  Subtract $d_{1}x^{\mu}g_{i}$ from $h$, add $d_{1}x^{\mu}$ to
  $f_{i}$ and return to Step 2. 
\end{enumerate}
\end{algo}
% \proof
% Let $h$ = $d_{1}t_{1}+\ldots+d_{m}t_{m}$
% We give proof of termination for border division of h where $ind_{\mathcal{O}}(h)$ = $1$ and 
% $d_{1}t_{1} \in \langle \partial\mathcal{O}_{s} \rangle_{A}$.
% \noindent
% Thus we can find $c_{\mu+1},\ldots,c_{s} \in A$ such that 
%     $c_{i}=0$ if $b_{i} \neq t_{1}$ and $d_{1}t_{1}=c_{\mu+1}b_{\mu+1}+\ldots+c_{s}b_{s}$.
% \endproof

% *Paragraph about the problem (Cyclic polynomials)*
% *Example showing the problem*
 
% To get around the above problem we define acyclic border prebasis
This procedure over rings differs from the case of
fields only in Step 4. 
%This is because of the notion of scalar
%border that exists in the first border.  
The termination of the above method is not assured because of the
possibility that  for a given polynomial, $f$, a monomial in its
support identified with index $0$ in Step 3 may again have an index 1
after Step 4. Therefore, we cannot assume the  reduction in index
values at every step of the procedure.  
%==========================================================
\section{Acyclic Border Prebases and Termination of Border Division Algorithm}\label{Acyclic Border PreBases}
\noindent
Here, we identify a special class of $\mathcal{O}$-border prebases
called acyclic  $\mathcal{O}$-border prebases for which  the
termination of the border division algorithm can be established.  
%From now onwards, we will be studying border bases for this class
%of $\mathcal{O}$-border prebases.

\begin{definition}\label{borderprebasisdefn}
A $\mathcal{O}$-border prebasis $G$ = $\{g_{1},\ldots,g_{s}\}$ is said
to be acyclic if there exists a permutation of $G$,
$\{g_{i_1},\ldots,g_{i_s}\}$ such that for any $g_{i_j}$, $g_{i_k}$,
where $j \lneq k$, exactly one of the following conditions are 
satisfied
\begin{enumerate}
 \item $c_{j}\op{BF}_{\mathcal{O}}(g_{i_j}) = c_{k}\op{BF}_{\mathcal{O}}(g_{i_k})$ for any $c_{j}, c_{k} \in A$ or
 \item $d_jx^{\alpha_j} \in \partial\mathcal{O}$ and
$d_jx^{\alpha_j} \in \op{supp}(g_{i_j})$ implies
$c_kx^{\alpha_j} \notin \op{supp}(g_{i_k})$ for some $c_k,d_j \in A$.
\end{enumerate}
\end{definition}
The ordered set of acyclic $\mathcal{O}$-border prebasis that satisfies the permutation given above is called a `well ordered' acyclic $\mathcal{O}$-border prebasis.
\noindent
We now show the correctness and termination of Algorithm~\ref{ProcedureDiv} when the $\mathcal{O}$-border prebasis is
acyclic.

\begin{proposition}\label{BorderDiv} {\bf (Border Division Algorithm)} 
Consider a polynomial $f \in A[x_1,\ldots,x_n]$. If the $\mathcal{O}$-border prebasis $G=\{g_{1},\ldots,g_{s}\}$ is acyclic, then Algorithm ~\ref{ProcedureDiv} terminates for $f$ and returns a tuple,  
\begin{displaymath}
(f_{1},\ldots,f_{s},a_{1},\ldots,a_{t}) \in (A[x_{1},\ldots,x_{n}])^{s} \times A^{t}
\end{displaymath}
 such that
\begin{displaymath}
 f=f_{1}g_{1}+\ldots+f_{s}g_{s}+a_{1}x^{\alpha_{1}}+\ldots+a_{t}x^{\alpha_{t}},
\end{displaymath}
and $\op{deg}(f_i) \leq \op{ind}_{\mathcal{O}}(f)$, for $i = 1,\ldots,s$ with 
$f_ig_i \neq 0$. 
\end{proposition}
\begin{proof}
We first describe the execution of the algorithm. In Step 4,
$\op{ind}_{\mathcal{O}}(d_1x^{\gamma_1}) = 1$ and $d_1x^{\gamma_1} 
\in \Span_A(\langle \partial\mathcal{O} \rangle_{A})$. This implies that $d_1 \in \mathcal{I}_{x^{\gamma_1}}$, where 
$\mathcal{I}_{x^{\gamma_1}}$ is an ideal generated by $\langle u_1,\ldots,u_k \rangle$, $\ u_ix^{\gamma_1} \in \partial\mathcal{O}_{s}$,
$1 \leq i \leq k$. Thus, there exists $l_1,\ldots,l_k \in A$ such that
$d_1= \sum_{i=1}^{k} l_iu_i$. Hence, $d_1x^{\gamma_1}= \sum_{s'+1}^{s} b_i(c_ix^{\beta_i})$,
where $c_ix^{\beta_i} \in \partial\mathcal{O}_{s}$ and
$b_i=l_j$ when  $c_ix^{\beta_i}=u_jx^{\gamma_1}$ for some  $j \in
\{1,\ldots,k\}$, and $b_i=0$, otherwise.
%\begin{displaymath}
%\begin{split}
%& c_i=a_j \text{ when } b_i=u_jx^{\beta_1} \text{ for some } j \in \{1,\ldots,\nu\},\\
%& c_i=0 \text{ otherwise.}
%\end{split}
%\end{displaymath}
The other steps, due to the absence of scalar border terms,
mimics the border basis division in fields \citep[Proposition 3]{KehreinKreuzer:2005:CharacterizationsOfBorderBases}. 
%\noindent
We prove that the representation, 
\begin{displaymath}
  f = f_{1}g_{1}+\ldots+f_{s}g_{s}+a_{1}x^{\alpha_{1}}+\ldots+a_tx^{\alpha_{t}}+h,
\end{displaymath}
computed by the algorithm is valid in every step. Clearly, it is satisfied in Step 1. In Step 4 we subtract
$(b_{s'+1}g_{s'+1}+\ldots+b_{s}g_{s})$ from  
$h$. These $b_i$s are then added to $f_i$s, i.e. $f_i=f_i+b_i$, $s'+1 \leq i \leq s$.
Similarly in Step 5, from $h$ we subtract $d_1x^{\mu}g_i$ and we add $d_1x^{\mu}$ to $f_i$. The constants $a_1,\ldots,a_t$
are modified only in Step 3. The representation of $f$ is valid because $\operatorname{ind_{\mathcal{O}}}(h)=0$.
If the algorithm terminates, $h=0$ and we have a valid representation. 

Now we prove that $\operatorname{deg}(f_i) \leq
\operatorname{ind}_{\mathcal{O}}(f)$ for all $i = 1,\ldots, s$. In Step 5 of the algorithm, where we divide using the 
monomial border, our choice of the term $d_1x^{\mu}$ is such that \\ $\operatorname{deg}(dx^{\mu})= \operatorname{ind}_{\mathcal{O}}(h)-1$. In Step 4, where we divide using the scalar border, the index of the intermediate 
polynomial, $h$ is 1. The $b_i$, $i = 1,\ldots, s$  are  constants and the degree of $f_i$, $i = 1,\ldots, s$ are therefore zero. All the other steps in the 
algorithm do not affect $f_i$, $i = 1,\ldots,s$. Also, in the algorithm the index of the intermediate polynomial, $h$ never increases.
From the above steps, the inequality
$\operatorname{deg}(f_i) \leq \operatorname{ind}_{\mathcal{O}}(f) - 1$ for all $i = 1,\ldots,s$ follows. 

Next, we prove that the algorithm terminates on all inputs.  
In Step 4, $\operatorname{ind_{\mathcal{O}}}(h)=1$ and
$\mathrm{Ter}(h) \subseteq$ $\Span_A( \Mon(\mathcal{O}) )
\cup \Span_A( \partial\mathcal{O}_{s} )$ = $\Span_A(
\Mon(\mathcal{O}) )$. We claim that Step 4 
terminates after a finite number of steps for an acyclic
$\mathcal{O}$-border prebasis. Let 
$h=d_1x^{\alpha_1}+\ldots+d_tx^{\alpha_t}$.
For simplicity, let us assume that the acyclic $\mathcal{O}$-border prebasis, $G$ is well ordered. It can easily be seen that $g_1$ will be
used atmost once in Step 4, while $g_2$ will be used at most twice ($h
\xrightarrow{g_2} h_1 \xrightarrow{G \setminus \{g_1,g_2\}}_{+} 
h_2 \xrightarrow{g_1} h_3 \xrightarrow{G \setminus \{g_1,g_2\}}_{+} h_4 \xrightarrow{g_2} h_5$). Similarly, any $g_i$ will be used atmost
$\operatorname{O}\bigl(i^2)$ times. For the set $G$, therefore Step 4 is executed at most $\operatorname{O}\bigl(s^3)$ times.
All the other steps of the division correspond to either order ideal, monomial border or the $k^{th}$ order border, $k \gneq 1$ and
therefore mimic the border division in fields. Hence, the termination
is guaranteed by \citep[Proposition
  3]{KehreinKreuzer:2005:CharacterizationsOfBorderBases}.  
\end{proof}
%Consider an ideal $\mathfrak{a} \subseteq A[x_1,\ldots,x_n]$ generated by an acyclic $\mathcal{O}$-border prebasis, 
%$G=\{g_1,\ldots,g_s\}$. 
The border division algorithm gives us the remainder upon division by an acyclic $\mathcal{O}$-border prebasis as a part of its output.
We now give a formal definition for remainder.

\begin{definition}\label{remainder}
  Let $\mathcal{O}$ be an order ideal and $\Mon(\mathcal{O})=\{x^{\alpha_{1}},\ldots,x^{\alpha_{t}}\}$,
  its monomial part. Let $G=\{g_{1},\ldots,g_{s}\}$ be the $\mathcal{O}$-border prebasis.
The $\mathcal{O}$-remainder of a polynomial $f$ w.r.t. $G$, if it exists, is given as 
 \begin{displaymath}
  \op{rem}_{\mathcal{O},G}(f)=a_{1}x^{\alpha_{1}}+\ldots+a_{t}x^{\alpha_{t}},
 \end{displaymath}
where $f=f_{1}g_{1}+\ldots+f_{s}g_{s}+a_{1}x^{\alpha_{1}}+\ldots+a_{t}x^{\alpha_{t}}$ and $a_i\in A$ for  all $i = 1,\ldots, t$ is a representation computed
by the border division algorithm whenever the algorithm terminates.
\end{definition}

%=======================================================
%\section{Some general forms of Macaulay Basis theorem}
\section{Order Span and Acyclic Border Bases}
\label{Acyclic Border Bases}
\noindent
%Let $\mathcal{O} \subseteq A \times \mathbb{N}^{n}$ be an order ideal
%and let $\Mon(\mathcal{O})$ be its monomial part. Given an ideal
%$\mathfrak{a}$, by $\Mon(\mathcal{O}) + \mathfrak{a}$ we mean  $
%\{x^{\alpha} + \mathfrak{a}  \ | \ x^{\alpha} \in \Mon(\mathcal{O})
%\}  $. 
Consider the case when $A[x_1,\ldots,x_n]/\mathfrak{a}$ is finitely generated. 
Using the order function we define a generating set  for
$A[x_1,\ldots,x_n]/\mathfrak{a}$ that also satisfies a weaker form of
the linear independence property. 
\begin{definition}[Order span]\label{weakstarbasis}
Let $\mathfrak{a}$ be an ideal and let $\mathcal{I}$ be a proper order function and $\mathcal{B}= \{x^{\alpha_1}, \ldots, x^{\alpha_m}\}$ be a finite set of monomials of size $m$ such that  
 $x^\alpha \notin \mathcal{B}$ if and only if $\mathcal{I}_{x^\alpha} = \langle 1 \rangle$, where $x^\alpha \in \mathbb{M}^{n}$. Let $C_{\mathcal{I}_{x^{\alpha}}}$ be the coset
representatives of the equivalence classes of $A/\mathcal{I}_{x^{\alpha}}$. Then we say the set of residue classes of $\mathcal{B}$ forms an order span for $ A[x_1,\ldots,x_n]/\mathfrak{a}$ 
w.r.t. $\mathcal{I}$ if it satisfies the following properties. 
\begin{enumerate}[(i)]
 \item $\mathcal{B} + \mathfrak{a}$ generates $ A[x_1,\ldots,x_n]/\mathfrak{a}$ as an $A$-module and 
\begin{displaymath}
 A[x_1,\ldots,x_n]/\mathfrak{a} =  \{ \sum\limits_{i=1}^{m} a_ix^{\alpha_i} + \mathfrak{a}\ |\ a_i \in 
 C_{\mathcal{I}_{x^{\alpha_i}}}, x^{\alpha_i} \in \mathcal{B} \}.
 \end{displaymath}
 \item If $\sum_{i=1}^{m} a_ix^{\alpha_i} + \mathfrak{a} = 0$, where $m \in \mathbb{N}$, $x^{\alpha_i} \in \mathcal{B}$ and $a_i \neq 0$ for all $i = 1, \ldots, m$, then for some 
 $j \in\{ 1,\ldots, m\}$, $a_j \in \mathcal{I}_{x^{\alpha_j}}$.
  \item If there exists an order function $\mathcal{I}^{'}$ such that   $\mathcal{I}^{'}_{x^\alpha} \subseteq \mathcal{I}_{x^\alpha}$ for some $x^\alpha \in \mathbb{M}^n$ and $\mathcal{B}^{'} = \{x^\alpha : \mathcal{I} ^{'}_{x^\alpha} \neq \langle 1 \rangle \}$ satisfies (i) and (ii) w.r.t. $\mathcal{I}^{'}$, then $\mathcal{I}^{'} = \mathcal{I}$. 
\end{enumerate}
\end{definition}
%Let $\mathfrak{a}$ be an ideal and let $\mathcal{B}$ be a
%set of monomials, possibly infinite, and let $\mathcal{I}$ be a coefficient ideal mapping such that $x^\alpha \notin
%\mathcal{B}$ if and only if  $\mathcal{I}_{x^\alpha} = \langle 1 \rangle$. We say that the set of residue classes
%of $\mathcal{B}$ forms a weak basis for $ A[x_1,\ldots,x_n]/\mathfrak{a}$ w.r.t. $\mathcal{I}$ if it satisfies the following properties. 
%\begin{enumerate}[(i)]
%\item $\mathcal{B} + \mathfrak{a}$ generates $ A[x_1,\ldots,x_n]/\mathfrak{a}$ as an $A$-module.
%\item If $\sum_{i=1}^{k} a_ix^{\alpha_i} + \mathfrak{a} = 0$, where $k \in \mathbb{N}$, $x^{\alpha_i}\in \mathcal{B}$ and $a_i \neq 0$ for all $i \in \{1, \ldots, k\}$, then for some 
% $j \in \{1,\ldots,k \}$, $a_j \in \mathcal{I}_{x^{\alpha_j}}$.
% \item If there exists a coefficient ideal mapping $\mathcal{I}^{'}$ such that   $\mathcal{I}^{'}_{x^\alpha} \subseteq \mathcal{I}_{x^\alpha}$ for some $x^\alpha \in \mathbb{M}^n$ and $\mathcal{B}^{'} = \{x^\alpha : \mathcal{I} ^{'}_{x^\alpha} \neq \langle 1 \rangle \}$ satisfies (i) and (ii) w.r.t. $\mathcal{I}^{'}$, then $\mathcal{I}^{'} = \mathcal{I}$. 
%\end{enumerate}
%\end{definition}
\begin{remark}
The linear independence property requires that if $\sum_{i=1}^{k}
a_ix^{\alpha_i} + \mathfrak{a} = 0$ then for all $i \in
\{1,\ldots,k\}$, $a_i \in \mathcal{I}_{x^{\alpha_i}}$. Therefore, the
second condition in Definition~\ref{weakstarbasis} is a weaker form of the
linear independence property.  In fact, in the case of fields and
residue class rings with a free $A$-module representation, the second condition  automatically
implies the linear independence property.   
\end{remark}
\begin{remark}
The third condition in Definition~\ref{weakstarbasis} can be interpreted
as a minimality condition on the spanning set.  Over fields, linear independence of the spanning set ensures minimality but over rings it has to be specified separately. Consider the ideal $\mathfrak{a} = \langle 4x_1, {x_1}^2,
x_2\rangle$ in  $\mathbb{Z}[x_1,x_2]$. Let $\mathcal{I}$ and
$\mathcal{I}^{'}$ be two order functions. Define
$\mathcal{I}_{1} = {0}, \mathcal{I}_{x_1} = \langle 2\rangle$ and for
all the other monomials $x^\alpha$,  $\mathcal{I}_{x^{\alpha}}
= \langle 1 \rangle$. Similarly define $\mathcal{I}^{'}_{1} = {0},
\mathcal{I}^{'}_{x_1} = \langle 4 \rangle$ and for all the other
$x^\alpha \in \mathbb{M}^2$,  $\mathcal{I}^{'}_{x^{\alpha}} = \langle
1 \rangle$. Both $\mathcal{I}$ and $\mathcal{I}^{'}$  satisfy the
first two conditions of the order span. However, since
$\mathcal{I}^{'}_{x_1}  (= \langle 4 \rangle) \subsetneq
\mathcal{I}_{x_1} ( = \langle 2 \rangle)$, it is w.r.t. the
second order function, $\mathcal{I}^{'}$ that we define the
order span of $\mathbb{Z}[x_1,x_2]/\langle 4x_1, {x_1}^2, x_2\rangle$.  
\end{remark}
%newest comments
%From the observations we have made above, it is easy to see that when
%the residue class ring, $A[x_1,\ldots,x_n]/ \mathfrak{a}$ is finitely
%generated  the structure satisfies the following additional
%properties.  
%
\begin{corollary}\label{weakbasiscorollary}
Let   $\mathfrak{a}$ be an ideal such
that  $A[x_1,\ldots,x_n]/ \mathfrak{a}$ is finitely generated.  Let
$G = \{g_1, \ldots, g_t\}$ be a Gr\"obner basis of an ideal $\mathfrak{a}$ and
$\mathcal{I}^{(G)}$ the order function fixed by $G$. Then,  
\begin{enumerate}
\item The order function $\mathcal{I}^{(G)}$ is proper,
%%\item The set of weak standard monomials w.r.t $\mathfrak{a}$ is
%  %finite,
%   and
\item The finite order span of $A[x_1,\ldots,x_n]/ \mathfrak{a}$ w.r.t. $\mathcal{I}^{(G)}$ is given by 
\begin{displaymath}
\mathcal{B} = \{ x^\alpha : \mathrm{lt}(g_i) \nmid ax^\alpha,  \text{ for some  nonzero } a \in A,  g_i \in G \}
\end{displaymath}
\end{enumerate}
\end{corollary}
We now provide a better interpretation of the mapping described by \eqref{equation} in terms 
of order span. 
%The proof of the following result can be found in Appendix~\ref{Proofappendix}.

\begin{proposition}\label{proposition1}
Let $\mathfrak{a}$ be an ideal in $A[x_1,\ldots,x_n]$. Let $\mathcal{I}$ 
be a proper order function and $\mathcal{B} = \{x^{\alpha_1},\ldots, x^{\alpha_m}\}$ be a finite set of monomials of size $m$ such that if $x^\alpha \notin \mathcal{B}$, 
$\mathcal{I}_{x^\alpha} = \langle 1 \rangle$, where $x^\alpha \in \mathbb{M}^{n}$. Consider a mapping, 
\[
\begin{split}
 \phi :  A[x_{1},\ldots,x_{n}]/\mathfrak{a} &\longrightarrow A/\mathcal{I}_{{x^{\alpha_1}}} \times \cdots \times A/\mathcal{I}_{{x^{\alpha_m}}}\\
         f + \mathfrak{a} &\longmapsto (c_1 +\mathcal{I}_{{x^{\alpha_1}}}  , \cdots, c_m+ \mathcal{I}_{x^{\alpha_m}}) ,
\end{split}
\]
where $c_i \in C_{\mathcal{I}_{x^{\alpha_i}}}$ for all $i \in \{1,\ldots, m\}$.
Then, $\phi$ is an isomorphism if $\mathcal{B}$ forms an order span for $A[x_1,\ldots,x_n]/\mathfrak{a}$. 
\end{proposition}
\begin{proof}

Let $\mathcal{B}$ form an order span for $A[x_1,\ldots,x_n]/\mathfrak{a}$. We first show that the 
 mapping $\phi$ is well defined. Consider a polynomial $f \in A[x_1,\ldots,x_n]$. Suppose 

 \begin{align*}
    \phi( f + \mathfrak{a}) & = (c_1 +\mathcal{I}_{{x^{\alpha_1}}}  , \cdots, c_m + \mathcal{I}_{x^{\alpha_m}}) \text{ and }
   \\  \phi(f + \mathfrak{a}) & =  (c'_1 +\mathcal{I}_{{x^{\alpha_1}}}  , \cdots, c'_m + \mathcal{I}_{x^{\alpha_m}}),
\end{align*}
\noindent where $c_i$, $c'_i$ $\in C_{\mathcal{I}_{x^{\alpha_i}}}$, for all $i = 1, \ldots, m$. This implies  
$(c_i-c'_i) \in \mathcal{I}_{{x^{\alpha_i}}}$, $i = 1, \ldots, m$. Since the difference of 
two different coset representatives cannot give the zero coset, we have $c_i = c'_i$ for all $i = 1, \ldots, m$. Thus $\phi$ is well defined.

\noindent
Clearly, $\phi$ is a surjective map by construction. We now have to prove that $\phi$ is an injective 
mapping. Consider a polynomial $f \in A[x_1,\ldots,x_n]$ such that $\phi(f+\mathfrak{a}) = (0,\ldots,0)$. Let us 
assume that $f \notin \mathfrak{a}$. Since $\mathcal{B}$ forms an order span for $A[x_1,\ldots,x_n]/\mathfrak{a}$,
we can obtain $c_i \in C_{\mathcal{I}_{x^{\alpha_i}}}$, $i = 1, \ldots, m$ such that $f = \sum\limits_{i=1}^{m} c_ix^{\alpha_i}$ mod 
$\mathfrak{a}$. Further, atleast one of the $c_i$, $i \in \{ 1, \ldots, m\}$, is nonzero. This implies that $\phi(f+ \mathfrak{a})$ 
also maps to $(c_1,\ldots, c_m)$. Therefore $\phi$ is not a well defined mapping. This is a 
contradiction and $f \in \mathfrak{a}$. Thus, the kernel of $\phi$, 
$\ker(\phi) = \{0 + \mathfrak{a}\}$. This implies that $\phi$ is an injective mapping. Hence, it follows that 
$\phi$ is an isomorphism. 
\end{proof}
\begin{definition}
\label{bbdefn}
Let $\mathfrak{a}$ be an ideal such that 
$A[x_{1},\ldots,x_{n}]/\mathfrak{a}$ is a finitely generated
$A$-module. Let $\mathcal{O}$ be an order ideal and
$G=\{g_{1},\ldots,g_{s}\}$ be an acyclic $\mathcal{O}$-border prebasis
consisting of polynomials in $\mathfrak{a}$. $G$ is an acyclic
$\mathcal{O}$-border basis of $\mathfrak{a}$ if $\Mon(\mathcal{O})$  is an order span of
$A[x_{1},\ldots,x_{n}]/ \mathfrak{a}$.  

% $A[x_{1},\ldots,x_{n}]/\mathfrak{a} = \{\sum a_ix^{\alpha_i}+\mathfrak{a}\ |\ x^{\alpha_i} \in %\mathcal{O}_{m} \text{ and } a_i \in A\}$ and $\sum a_ix^{\alpha_i} +
%\mathfrak{a} = 0$ implies $a_i \in \mathcal{I}_{x^{\alpha_i}}$, for some $i$.
\end{definition}
%Note that since the coefficient ideal mapping is proper and the
%corresponding monomial part of $\mathcal{O}$, $\Mon(\mathcal{O})$ is
%finite we consider only 
%a finitely generated $A$-module,
%$A[x_1,\ldots,x_n]/\mathfrak{a}$. This is analogous to the study of
%border bases over fields since  
%there we study the finitely generated $\Bbbk$-vector space,
%$\Bbbk[x_1,\ldots,x_n]/\mathfrak{a}$. 
The next proposition shows that these polynomials
indeed generate the ideal in $\mathfrak{a}$. 
%The proof can be found in Appendix~\ref{Proofappendix}. 
\begin{proposition}\label{idealgen} 
Let $\mathfrak{a}$ be an ideal such
that $A[x_{1},\ldots,x_{n}]/\mathfrak{a}$ is finitely generated as an
$A$-module. Let $\mathcal{O}$ be an order ideal and let
$G=\{g_{1},\ldots,g_{s}\}$ be an acyclic $\mathcal{O}$-border basis
for $\mathfrak{a}$. Then $\mathfrak{a}$ is generated by $G$. 
\end{proposition}
\begin{proof} 
Let $\Mon(\mathcal{O})=\{x^{\alpha_{1}},\ldots,x^{\alpha_{t}}\}$ be the monomial part of $\mathcal{O}$ and
$G=\{g_{1},\ldots,g_{s}\}$ be an acyclic $\mathcal{O}$-border basis of $\mathfrak{a}$. Consider, $f\in \mathfrak{a}$.
By Algorithm \ref{BorderDiv}, we have $f_{1},\ldots,f_{s} \in A[x_{1},\ldots,x_{n}]$ and 
$a_{1},\ldots,a_{t} \in A$ such that
\begin{equation}\label{eq1}
f=f_{1}g_{1}+\ldots+f_{s}g_{s}+a_{1}x^{\alpha_{1}}+\ldots+a_{t}x^{\alpha_{t}}.
\end{equation}
Now,
$f-\sum_{i=1}^{s}f_{i}g_{i} \in \mathfrak{a}$.
This implies,
$\sum_{i=1}^{t}a_{i}x^{\alpha_{i}} \in \mathfrak{a}$. 
 Let $h = \sum_{i=1}^{t}a_{i}x^{\alpha_{i}}$. 
Suppose $h \neq 0$, then $a_{i} \notin \mathcal{I}_{x^{\alpha_{i}}} \setminus \{0\}$, for all $i = 1,\ldots,t$. 
If $a_{i} \in \mathcal{I}_{x^{\alpha_{i}}}$, for all $i = 1,\ldots,t$,  then $\operatorname{ind_{\mathcal{O}}}(h) = 1$. 
Then, \eqref{eq1} is not a valid output of Algorithm ~\ref{BorderDiv}. But we are also given that $G$ is an acyclic
border basis of $\mathfrak{a}$. The order span property of border basis implies that if
$\sum_{i=1}^{t}a_{i}x^{\alpha_{i}} \in \mathfrak{a}$ for $a_i \neq 0$ for all $i = 1,\ldots,t$, then
$a_i \in \mathcal{I}_{x^{\alpha_{i}}}$ for some $i \in \{1,\ldots,t\}$. This is a contradiction. Hence, $h = 0$.
 We have, $f =  \sum_{i=1}^{s}f_{i}g_{i}$.
The other inclusion follows from the fact that $G \subseteq \mathfrak{a}$. 
\end{proof}
We need to verify if an acyclic $\mathcal{O}$-border basis
exists for every ideal, $\mathfrak{a}$ in $A[x_1,\ldots,x_n]$.  We
first address, below, whether  an acylic
$\mathcal{O}$-border basis for $\mathfrak{a}$ given an order ideal,
$\mathcal{O}$, exists. We also prove the uniqueness of the acyclic
$\mathcal{O}$-border basis.  
%The proof can be found in Appendix~\ref{Proofappendix}. 

\begin{theorem}\label{UniqueBB}
Let $\mathcal{O}$ be an order ideal, and let $\mathfrak{a}$ be an
ideal such that   
 $A[x_1,\ldots,x_n]/\mathfrak{a}$ is a finitely generated $A$-module.
If $\Mon(\mathcal{O})$ is an order span
% Suppose $A[x_{1},\ldots,x_{n}]/\mathfrak{a}= \{ \sum a_ix^{\alpha_i}+\mathfrak{a}\ |\ %x^{\alpha_i} \in \mathcal{O}_{m} \text{ and }
 %a_i \in A\}$ and $\sum a_ix^{\alpha_i} + \mathfrak{a} = 0$ implies for some $i$, $a_i \in %\mathcal{I}_{x^{\alpha_i}}$,
then there exists a unique acyclic $\mathcal{O}$-border basis for
$\mathfrak{a}$. 
\end{theorem}
\begin{proof}
Let $\Mon(\mathcal{O})=\{x^{\alpha_{1}},\ldots,x^{\alpha_{t}}\}$ be
the monomial part of $\mathcal{O}$, and let
$\partial\mathcal{O}=\{c_{1}x^{\beta_1},\ldots,c_{s}x^{\beta_s}\}$ be
the border of $\mathcal{O}$. We now prove that an $\mathcal{O}$-border
basis exists for $\mathfrak{a}$. Since $\Mon(\mathcal{O})$ is an order span basis,
for each $c_ix^{\beta_i} \in \partial\mathcal{O}$ one can find
$a_{j}^{(i)}x^{\alpha_i}s \in \mathcal{O}$  such that $c_ix^{\beta_i} = \sum_{j=1}^{t}
a^{(i)}_{j}x^{\alpha_j} \text{ mod } \mathfrak{a}$. 
% This is because % $c_{i}x^{\beta_{i}} -
% \sum_{j=1}^{t} a_{ij}x^{\alpha_{j}} \in
% \mathfrak{a}$ and hence $c_i-a_{ij} \in
% \mathcal{I}_{x^{\alpha_{j}}}$.  
% This can happen only when $a_{ij}=0$.
We define $G$ as
\[
G= \{c_ix^{\beta_{i}} - \sum_{j=1}^{t} a^{(i)}_{j}x^{\alpha_{j}}\ |\ 1 \leq i \leq s\}.
\]
Clearly, $G$ is an acyclic $\mathcal{O}$-border prebasis. Now, $G \subseteq \mathfrak{a}$ and $A[x_{1},\ldots,x_{n}]/\mathfrak{a}=
\{\sum a_ix^{\alpha_i}+\mathfrak{a}\ |\ x^{\alpha_i} \in \Mon(\mathcal{O}) \text{ and } a_i \in A\}$. Hence, $G$ is an $\mathcal{O}$-border 
basis of $\mathfrak{a}$.

To prove the uniqueness of $\mathcal{O}$-border basis, consider two acyclic $\mathcal{O}$-border bases
for $\mathfrak{a}$. Let $G=\{g_{1},\ldots,g_{t}\}$ and
$G=\{g'_{1},\ldots,g'_{t}\}$ such that
\begin{displaymath}
 \begin{split}
 & g_{i} = c_{i}x^{\beta_i} - \sum_{j=1}^{t} a^{(i)}_{j}x^{\alpha_{j}}, \text{ where each } a_{j}^{(i)}x^{\alpha_{j}} \in \mathcal{O} \text{ and},\\
 & g'_{i}= c_{i}x^{\beta_i}-\sum_{j=1}^{t} a'^{(i)}_{j}x^{\alpha_{j}}, \text{ where each } a'^{(i)}_{j}x^{\alpha_{j}} \in \mathcal{O}.\\
\end{split}
\end{displaymath}
We have,
\begin{displaymath}
  g_{i}-g'_{i}= \sum_{j=1}^{t}
  a^{(i)}_{j}x^{\alpha_{j}}-\sum_{j=1}^{t} a'^{(i)}_{j}x^{\alpha_{j}}
  \in \mathfrak{a}.\\ 
\end{displaymath}
This implies,
\begin{displaymath}
\sum_{j=1}^{t} (a^{(i)}_{j}-a'^{(i)}_{j})x^{\alpha_{j}} \in \mathfrak{a}. 
\end{displaymath}
Since $a^{(i)}_{j}$ and $a'^{(i)}_j$ are coset representatives of
$A/\mathcal{I}_{x^{\alpha_j}}$ and the difference of two different cosets
cannot be a zero coset, we have
$a^{(i)}_j=a'^{(i)}_j$. Therefore, $g_{i}=g'_{i}$. Hence, the acyclic
$\mathcal{O}$-border basis of $\mathfrak{a}$ is unique. 
\end{proof}
Thus, the question of existence of a border basis for an ideal reduces
to the following questions. Given an ideal $\mathfrak{a}$, (i) does there
always exist a proper order function, $\mathcal{I}$ such that the monomial part of the order
ideal, $\Mon(\mathcal{O})$ constructed from $\mathcal{I}$ forms an order span
for $A[x_{1},\ldots,x_{n}]/\mathfrak{a}$  and (ii) will the 
corresponding $\mathcal{O}$-border basis be acyclic.  
We use the theory of Gr\"obner bases to establish the result.
\begin{theorem}\label{BBexist}
 Given an ideal $\mathfrak{a}$ such that $A[x_{1},\ldots,x_{n}]/\mathfrak{a}$  is finitely generated as an $A$-module, 
 there always exists an acyclic border basis of $\mathfrak{a}$ corresponding to 
 some order ideal, $\mathcal{O}$. 
%  if  $\Mon(\mathcal{O}) + \mathfrak{a}$ is a weak basis. 
 % $A[x_{1},\ldots,x_{n}]/\mathfrak{a}$ = $\{\sum a_ix^{\alpha_i}+\mathfrak{a} \ |\ x^{\alpha_i} %\in \mathcal{O}_{m} \text{ and } 
% a_i \in A\}$ and $\sum a_ix^{\alpha_i} + \mathfrak{a} = 0$ implies for some $i$, $a_i \in %\mathcal{I}_{x^{\alpha_i}}$.
\end{theorem}
\begin{proof} Let $\prec$ be a monomial order on $A[x_{1},\ldots,x_{n}]$. Let $G'=\{g'_{1},\ldots,g'_{t}\}$ be a Gr\"obner basis 
of $\mathfrak{a}$. Consider the order function fixed by $G$, $\mathcal{I}^{(G)}$. Since $A[x_{1},\ldots,x_{n}]/\mathfrak{a}$  is finitely generated, the mapping is 
proper. 
Let  $\mathcal{O}_{\prec}$ be the order ideal corresponding to $\mathcal{I}^{(G)}$. 
%Commenting this sentence. Nithish needs to confirm. 
It follows from Corollary~\ref{weakbasiscorollary} that
%Clearly,
$\Mon(\mathcal{O}_{\prec})$ forms an order span. 
% \[
%   \mathcal{O}_{\prec} = \{ax^{\alpha}\ |\ a \in C_{J_{x^{\alpha}}} \text{ where } (I_{J_{x^{\alpha}}},x^{\alpha}) \in B_{\prec}\}.
% \]
% From Equation ~\ref{equation}, we have if 
% $A[x_{1},\ldots,x_{n}]/\mathfrak{a}$ is a finitely generated $A$-module of size $t$, 
% then there exists an isomorphism $\phi$ between $A[x_{1},\ldots,x_{n}]/\mathfrak{a}$ 
% and $A/I_{J_{x^{\alpha{1}}}} \times \cdots \times A/I_{J_{x^{\alpha_{t}}}}$  defined as
% \begin{align*}
%  \phi :  A[x_{1},\ldots,x_{n}]/\mathfrak{a} &\longrightarrow A/I_{J_{x^{\alpha_1}}} \times \cdots \times A/I_{J_{x^{\alpha_t}}}\\
%          \sum\limits_{i=1}^t a_i x^{\alpha_i} + \mathfrak{a} &\longmapsto (c_1 + \mathcal{I}_{x^{\alpha_{1}}}, \cdots, c_t+ \mathcal{I}_{x^{\alpha_{t}}}),   
% \end{align*}
% where $c_i = a_i \text{  mod  } I_{J_{x^{\alpha_i}}}$ and  $c_i \in C_{J_{x^{\alpha_i}}}$.Details of the construction of this isomorphism can be found in \citep{FrancisDukkipati:2014:OnReducedGrobnerBasisAndMBtheorem}. Since the number of monomials is finite, 
% the set $B_{\prec}$ is finite and $A[x_{1},\ldots,x_{n}]/\mathfrak{a}$ = $\{\sum a_ix^{\alpha_i}+\mathfrak{a} \ |\ x^{\alpha_i} 
% \in \Mon(\mathcal{O}_\prec) \text{ and }  a_i \in A\}$. 
% Let $\Mon({\mathcal{O}_{\prec}})=\{x^{\alpha_1},\ldots,x^{\alpha_t}\}$ be the monomial part of $\mathcal{O}_\prec$.
Let $\partial\mathcal{O_{\prec}}=
 \{c_{1}x^{\beta_{1}},\ldots,c_{s}x^{\beta_s}\}$ be the border of $\mathcal{O}_{\prec}$. 
Let $G$ be the $\mathcal{O}_{\prec}$-border prebasis constructed along the same lines as in the proof of Theorem ~\ref{UniqueBB}. 
Therefore, each polynomial $g_i$ in $G$ is of the form, 
\begin{displaymath}
g_i= c_{i}x^{\beta_i} - \sum_{j=1}^{t} a^{(i)}_{j}x^{\alpha_{j}}, \hspace{3pt} i = 1, \ldots, s,
\end{displaymath}
where $c_ix^{\beta_i} \in \partial\mathcal{O}$ and each
$a^{(i)}_{j}x^{\alpha_{j}} \in \mathcal{O}$.   

For any $g_{i} \in G$, monomial ordering imposes that for every nonzero 
$a^{(i)}_{j}x^{\alpha_j}$,  $x^{\alpha_{j}} \prec x^{\beta_i}$. Also,
for two distinct border terms $c_{i}x^{\beta_{i}}$ and
$c_{j}x^{\beta_{j}}$  such that $x^{\beta_{i}} \neq x^{\beta_{j}}$,
either $x^{\beta_{i}} \prec x^{\beta_{j}} \text{ or } x^{\beta_{j}}
\prec  x^{\beta_{i}}$. The acyclic property of $G$ follows from these
two observations. Theorem \ref{UniqueBB} implies that $G$ forms a
unique acyclic $\mathcal{O_{\prec}}$-border basis for $\mathfrak{a}$.  
\end{proof}
For any polynomial $f \in A[x_1,\ldots,x_n]$, given an acylic $\mathcal{O}$-border prebasis, $G$ and an order ideal, $\mathcal{O}$, $\mathcal{O}$-remainder of $f$ is denoted by $ \operatorname{rem_{\mathcal{O},G}}(f)$ (Definition~\ref{remainder}).
\begin{proposition}\label{IdealRem} 
 Let $\mathfrak{a} \subseteq A[x_{1},\ldots,x_{n}]$ be an ideal such that $A[x_{1},\ldots,x_{n}]/\mathfrak{a}$ is 
 finitely generated as an $A$-module and $G \subseteq \mathfrak{a}$ be
 an acyclic $\mathcal{O}$-border basis of 
 $\mathfrak{a}$. For any $f \in A[x_{1},\ldots,x_{n}]$, $f \in
 \mathfrak{a}$ if and only if  
 $\mathrm{rem}_{\mathcal{O},G} (f) = 0$. 
\end{proposition}
\begin{proof}
Let $\Mon(\mathcal{O})=\{x^{\alpha_{1}},\ldots,x^{\alpha_{t}}\}$ be the monomial part of $\mathcal{O}$ and
$G=\{g_{1},\ldots,g_{s}\}$ be an acyclic $\mathcal{O}$-border basis of $\mathfrak{a}$. 
By Algorithm \ref{BorderDiv}, we have $f_{1},\ldots,f_{s} \in A[x_{1},\ldots,x_{n}]$ and 
$a_{1},\ldots,a_{t} \in A$ such that
\begin{equation}\label{eq2}
 f= \sum_{i=1}^{s} f_{i}g_{i} + \mathrm{rem}_{\mathcal{O,G}} (f),
\end{equation}
where $\operatorname{rem}_{\mathcal{O},G} (f) = \sum_{j=1}^{t} a_{j}x^{\alpha_{j}}$.
If $\mathrm{rem}_{\mathcal{O},G} (f)=0$ then $f=\sum_{i=1}^{s}
f_{i}g_{i}$. Hence $f \in \mathfrak{a}$.
 
Now let $f \in \mathfrak{a}$. Then $f-\sum_{i=1}^{s} f_{i}g_{i} \in \mathfrak{a}$. This implies, $\operatorname{rem}_{\mathcal{O},G} (f) 
\in \mathfrak{a}$. Suppose $\operatorname{rem}_{\mathcal{O},G} (f)$ is
not equal to zero. The proof proceeds along the same lines as the
proof of Proposition~\ref{idealgen}, where we take  $h =
\operatorname{rem}_{\mathcal{O},G} (f)$ and arrive at a
contradiction. Hence, when $f \in \mathfrak{a}$ the remainder  of $f$
w.r.t. $G$ is zero. 
\end{proof}
%\begin{proof}
%Let $\Mon(\mathcal{O})=\{x^{\alpha_{1}},\ldots,x^{\alpha_{t}}\}$ be the monomial part of $\mathcal{O}$ and
%$G=\{g_{1},\ldots,g_{s}\}$ be an acyclic $\mathcal{O}$-border basis of $\mathfrak{a}$. 
%By Algorithm \ref{BorderDiv}, we have $f_{1},\ldots,f_{s} \in A[x_{1},\ldots,x_{n}]$ and 
%$a_{1},\ldots,a_{t} \in A$ such that
%\begin{equation}\label{eq2}
% f= \sum_{i=1}^{s} f_{i}g_{i} + \mathrm{rem}_{\mathcal{O,G}} (f),
%\end{equation}
%where $\operatorname{rem}_{\mathcal{O},G} (f) = \sum_{j=1}^{t} a_{j}x^{\alpha_{j}}$.
%If $\mathrm{rem}_{\mathcal{O},G} (f)=0$ then $f=\sum_{i=1}^{s}
%f_{i}g_{i}$. Hence $f \in \mathfrak{a}$.
% 
%Now let $f \in \mathfrak{a}$. Then $f-\sum_{i=1}^{s} f_{i}g_{i} \in \mathfrak{a}$. This implies, $\operatorname{rem}_{\mathcal{O},G} (f) 
%\in \mathfrak{a}$. Suppose $\operatorname{rem}_{\mathcal{O},G} (f)$ is
%not equal to zero. The proof proceeds along the same lines as the
%proof of Proposition~\ref{idealgen}, where we take  $h =
%\operatorname{rem}_{\mathcal{O},G} (f)$ and arrive at a
%contradiction. Hence, when $f \in \mathfrak{a}$ the remainder  of $f$
%w.r.t. $G$ is zero. 
%\end{proof}

The above proposition  enables us to solve the ideal membership problem provided the acyclic border basis of the ideal is known.
However, it must be noted that the remainder on division by an acyclic  $\mathcal{O}$-border basis 
 for any $f \in A[x_{1},\ldots,x_{n}]$ is not unique. 

Below we define the normal form of a polynomial w.r.t. an acyclic border basis.
\begin{definition}
Let $\mathfrak{a}$ be an ideal such that
$A[x_{1},\ldots,x_{n}]/\mathfrak{a}$ is   finitely generated as an
$A$-module. Let $G \subseteq \mathfrak{a}$ be an acyclic
$\mathcal{O}$-border basis for   $\mathfrak{a}$. Let
$\Mon(\mathcal{O})=\{x^{\alpha_1},\ldots,x^{\alpha_t}\}$ be the
monomial part of $\mathcal{O}$, $C_{\mathcal{I}_{x^{\alpha_i}}}$ be the set of coset representatives
of the equivalence classes $A/\mathcal{I}_{x^{\alpha_i}}$ and $f$ be any polynomial in
$A[x_1,\ldots,x_n]$. Let $r$ be a polynomial given by
$r=a_1x^{\alpha_1}+\cdots+a_tx^{\alpha_t}$, where $a_i \in C_{\mathcal{I}_{x^{\alpha_i}}}$, $1 \leq i \leq
t$. Then $r$ is said to be the normal form of $f$ if $f = r \text{ mod } \mathfrak{a}$. 
\end{definition}

The normal form of a polynomial is denoted by
$\operatorname{NF}_{\mathcal{O},G}(f)$. We now prove that every
polynomial $f$ in  $A[x_1,\ldots,x_n]$ has a unique normal form. 

\begin{proposition}\label{uniquenormalform}
Let $\mathfrak{a} \subseteq A[x_{1},\ldots,x_{n}]$ be an ideal such
that $A[x_{1},\ldots,x_{n}]/\mathfrak{a}$ is  finitely generated as an
$A$-module. Let $G \subseteq \mathfrak{a}$ be an acyclic
$\mathcal{O}$-border basis for   $\mathfrak{a}$. For any polynomial
$f$ in $A[x_1,\ldots,x_n]$, the normal form of $f$ is unique.  
\end{proposition}
\begin{proof}
Let $\Mon(\mathcal{O})=\{x^{\alpha_{1}},\ldots,x^{\alpha_{t}}\}$ be the monomial part of $\mathcal{O}$. Let $s' \leq t$ be the number
of monomials in the scalar border, $\partial\mathcal{O}_s$ and $G=\{g_{1},\ldots,g_{s}\}$ be an acyclic $\mathcal{O}$-border basis 
of $\mathfrak{a}$. The existence of a normal form, $\operatorname{NF}_{\mathcal{O},G}(f)$
for any polynomial $f \in A[x_1,\ldots,x_n]$ is a consequence of the following equality,
\[
 A[x_1,\ldots,x_n]/\mathfrak{a} =  \{ \sum\limits_{i=1}^{k} a_ix^{\alpha_i} + \mathfrak{a}\ |\ a_i \in 
 C_{\mathcal{I}_{x^{\alpha_i}}} \} . 
\]

\noindent 
Now, we prove the uniqueness of the normal form of $f$. 
Let $r_1$ and $r_2$ be two different normal forms for $f$. Then $f = r_1 \text{ mod } \mathfrak{a}$ and $f = r_2 \text{ mod } \mathfrak{a}$.
This implies, $r_1 = r_2 \text{ mod } \mathfrak{a}$. Therefore, $r_1-r_2 \in \mathfrak{a}$. Let
$r_1=\sum_{i=1}^{t} b_{i}x^{\alpha_i}$ and 
$r_2=\sum_{i=1}^{t} b'_{i}x^{\alpha_i}$, where $b_i$ and $b'_i$ are
coset representatives in $C_{\mathcal{I}_{x^{\alpha_i}}}$. Then, 
 $r_1-r_2=\sum_{i=1}^{t} (b_i-b'_i)x^{\alpha_i}$.
If $r_1 \neq r_2$, then there is atleast one $i$ such that $b_i \neq b'_i$. This implies that $b_i-b'_i \neq 0$. Since $b_i$ and
$b'_i$ are coset representatives of distinct cosets, $b_i - b'_i \notin \mathcal{I}_{x^{\alpha_i}}$. 
Therefore, $(r_1-r_2) \notin \mathfrak{a}$. Hence a contradiction. Thus $r_1=r_2$ and the normal 
form of a polynomial is unique.
\end{proof}

The below result states that if we can associate a monomial order to an order ideal $\mathcal{O}$, then the reduced Gr\"obner basis of $\mathfrak{a}$ w.r.t. to that 
monomial order is a subset of the acyclic border basis associated with $\mathcal{O}$. 
%Appendix~\ref{Proofappendix} contains the proof details. 
\begin{proposition}\label{Pauerinborder}
Let $\mathfrak{a}$ be an ideal. Let $\prec$ be a
monomial order. Let $\mathcal{O}_{\prec}$  be an order ideal
corresponding to $\prec$ such that $\Mon(\mathcal{O})$ forms an
order span for $A[x_{1},\ldots,x_{n}]/\mathfrak{a}$. Then
Pauer reduced
Gr\"obner basis \citep{Pauer:2007:GrobnerBasesWithCoefficientsInRings} of $\mathfrak{a}$ w.r.t. $\prec$ is a subset of the
acyclic $\mathcal{O}$-border basis of $\mathfrak{a}$.   
\end{proposition}
\begin{proof}
Let $\Mon(\mathcal{O}_{\prec}) = \{x^{\alpha_1},\ldots,x^{\alpha_t}\}$
be the monomial part of  $\mathcal{O}_{\prec}$ and
$\partial\mathcal{O}_{\prec} =
\{c_1x^{\beta_1},\ldots,c_sx^{\beta_s}\}$ be the border of
$\mathcal{O}_{\prec}$. Let $G=\{g_1,\ldots,g_s\}$ be the acyclic
$\mathcal{O}_{\prec}$-border basis for $\mathfrak{a}$. The acyclic
$\mathcal{O}_{\prec}$-border basis is constructed as in the proof of
Theorem~\ref{BBexist}. Since $G$ is an acyclic
$\mathcal{O}_{\prec}$-border basis we have from
Proposition~\ref{IdealRem} that for any $f \in \mathfrak{a}$, $f$
reduces to zero. 
This implies that $G$ is a Gr\"obner basis of $\mathfrak{a}$. Further,
$\langle \operatorname{Lt}(\mathfrak{a}) \rangle$ is generated by  
$\partial\mathcal{O}_{\prec}$.
%We consider the reduced Gr\"obner basis
%over rings as defined in
%\citep{Pauer:2007:GrobnerBasesWithCoefficientsInRings}. 
Recall that, $\langle \operatorname{Lc}(\alpha,\mathfrak{a}) \rangle = \langle \operatorname{Lc}(f) :f \in \mathfrak{a},\ 
\operatorname{Lm}(f)=x^{\alpha} \rangle$ and $\langle \operatorname{Lc}(<\alpha,\mathfrak{a}) \rangle = \langle 
\operatorname{Lc}(f) : f \in \mathfrak{a},\ \alpha \in \operatorname{deg}(f)+ \mathbb{M}^{n},\ 
\operatorname{Lm}(f) \neq x^{\alpha} \rangle$. Clearly, for each monomial $x^{\alpha}$, in the monomial part, 
$\Mon(\mathcal{O}_{\prec})$, $\langle \operatorname{Lc}(\alpha,\mathfrak{a}) \rangle = \mathcal{I}_{x^{\alpha}}$. From the definition of
order ideal $\langle \operatorname{Lc}(\alpha,\mathfrak{a}) \rangle \neq \langle 1 \rangle$. For each
monomial $x^{\alpha}$ in the monomial border,  $\partial\mathcal{O}_{\prec_{m}}$, 
$\langle \operatorname{Lc}(\alpha,\mathfrak{a}) \rangle = \langle 1 \rangle$. Also, for each monomial $x^{\alpha} \in \operatorname{Lm}(G)$, we
have
\begin{displaymath}
 \mathrm{Gen} (\alpha, \mathfrak{a}) = \{\eta(a,\langle\mathrm{Lc}(<\alpha,\mathfrak{a})\rangle) : a \in \mathrm{Gen}(\langle\mathrm{Lc}(\alpha,  \mathfrak{a})\rangle)\} \setminus\{0\},
 \end{displaymath}
where $\eta(a,\langle\mathrm{Lc}(<\alpha,\mathfrak{a})\rangle) $ maps to an element in the coset $a + \langle\mathrm{Lc}(<\alpha,\mathfrak{a})\rangle$.
Consider the set, $\partial\mathcal{O}_{\prec_{\mathrm{red}}} =
\{cx^{\alpha} \in \partial\mathcal{O}_{\prec} : c \notin \langle \operatorname{Lc}(<\alpha,\mathfrak{a})  
\rangle\}$. This set
contains all the terms of the form $cx^{\alpha}$ in the border
$\partial\mathcal{O}_{\prec}$ such that $c$ cannot be expressed as a
combination of leading coefficients of those monomials that properly  divide
$x^{\alpha}$. Clearly, $\partial\mathcal{O}_{\prec_{\mathrm{red}}}
\subseteq \partial\mathcal{O}_{\prec}$. Let   $G_{\mathrm{red}}
\subseteq G$ consist of polynomials in $G$ with the border term in
$\partial\mathcal{O}_{\prec_{\mathrm{red}}}$. It can easily be seen
that $\langle \operatorname{Lt}(\mathfrak{a}) \rangle= \langle
\partial\mathcal{O}_{\prec_{\mathrm{red}}} \rangle$. Therefore,   
$G_{\mathrm{red}}$ is a Gr\"obner basis for $\mathfrak{a}$. Also, it
is clear from the construction of
$\partial\mathcal{O}_{\prec_{\mathrm{red}}}$ that
$\operatorname{Gen}(\alpha,\mathfrak{a})=\{c\ :\ cx^{\alpha} \in
\partial\mathcal{O}_{\prec_{\mathrm{red}}}\}$. We now prove that $G$ 
satisfies the two properties of Pauer's reduced Gr\"obner basis. 
The bijectivity of the map,
 \begin{align*}
   \phi: \{g\in G_{\mathrm{red}} : \mathrm{deg}(g) = \alpha \}
   &\longrightarrow \op{Gen} (\alpha,\mathfrak{a})\\ 
 \phi(g)&\longmapsto \op{Lc}(g) 
\end{align*}
follows from the observation that corresponding to each border term
$cx^{\beta} \in \partial\mathcal{O}_{\prec}$, there is  
exactly one polynomial $g \in G$ such that the border term in $g$ is $cx^{\beta}$. If we had not considered the reduced border
$\partial\mathcal{O}_{\prec_{\mathrm{red}}}$, then for all $g \in G \setminus G_{\mathrm{red}}$, $\phi(g)$ will map to zero. Also, each 
polynomial $g_i \in G_{\mathrm{red}}$ is of 
the form, 
\[
c_ix^{\beta_i}-\sum_{j=1}^{t} a^{(i)}_{j}x^{\alpha_j} \text{ where } c_ix^{\beta_i} \in \partial\mathcal{O}_{\prec} \text{ and } 
a^{(i)}_{j}x^{\alpha_j} \in \mathcal{O}_{\prec}.
\]
Since for each $a^{(i)}_{j}x^{\alpha_j}$, $a^{(i)}_{j} \in A/\mathcal{I}_{x^{\alpha_j}}$ we have that   \\
$\eta(a^{(i)}_{j},\langle\mathrm{Lc}(\alpha_j,\mathfrak{a})\rangle)=a^{(i)}_{j}$. Hence
$G_{\mathrm{red}}$ satisfies the second condition of Pauer's reduced
Gr\"obner basis. Therefore, $G_{\mathrm{red}}$ is a reduced Gr\"obner
basis for $\mathfrak{a}$. 
\end{proof}

\begin{theorem}\label{BBequiv}
 Let $\mathfrak{a}$ be a nonzero ideal in $A[x_{1},\ldots,x_{n}]$, $\mathcal{O}$ be an order ideal and
 $\partial\mathcal{O}=\{c_{1}x^{\beta_1},\ldots,c_{s}x^{\beta_s}\}$ be its border. Let $\Mon(\mathcal{O})=\{x^{\alpha_1},\ldots,x^{\alpha_t}\}$ 
 be the monomial part of $\mathcal{O}$ and $G =\{g_1, \ldots, g_s\}$ be an acyclic $\mathcal{O}$-border
 prebasis. Then the following statements are
 equivalent.
\begin{enumerate}
\item[(i)] $G$ is an acyclic $\mathcal{O}$-border basis for $\mathfrak{a}$.
\item[(ii)] $f \in \mathfrak{a}$ if and only if $f \xrightarrow{G}_{+}$ 0.
%\item[(ii)]  $\operatorname{BF}_{\mathcal{O}}(I) = \langle c_{1}x^{\beta_1},\ldots,c_{s}x^{\beta_s} \rangle$.
\item[(iii)] $f \in \mathfrak{a}$ if and only if there exists $f_1,\ldots,f_s \in A[x_1,,\ldots,x_n]$ such that $f=\sum_{i=1}^{s} f_ig_i$ and $\operatorname{max}\{ \operatorname{deg}(f_i)\ 
  |\ f_ig_i \neq 0, \hspace{1pt} i = 1,\ldots,s\} = \operatorname{ind_{\mathcal{O}}}(f)-1$.
\item[(iv)] The border form of $\mathfrak{a}$, $\operatorname{BF}_{\mathcal{O}}(\mathfrak{a}) = \langle c_1x^{\beta_1}, \ldots, c_sx^{\beta_s} \rangle$.
\end{enumerate}
\end{theorem}
\begin{proof}
(i) $\Rightarrow$ (ii). The claim follows from the proof of Proposition~\ref{IdealRem}.

\noindent
(ii) $\Rightarrow$ (iii). Let $f \in \mathfrak{a}$. By the border division algorithm, there exists $f_1,\ldots,f_s \in A[x_1,,\ldots,x_n]$, 
$\operatorname{deg}(f_i) \leq \operatorname{ind_{\mathcal{O}}}(f)-1$, $1 \leq i \leq s$ such that $f=\sum_{i=1}^{s} f_ig_i$. Assume that 
$\operatorname{max}\{ \operatorname{deg}(f_i)\} \lneq \operatorname{ind_{\mathcal{O}}}(f)-1$. It can easily be verified that 
$\operatorname{ind_{\mathcal{O}}}(f_ig_i) \leq \operatorname{ind_{\mathcal{O}}}(g_i)+\operatorname{deg}(f_i)$. The definition of
$\mathcal{O}$-border prebasis implies that $\operatorname{ind_{\mathcal{O}}}(g_i)=1$. Thus
$\operatorname{ind_{\mathcal{O}}}(f_ig_i) \leq \operatorname{deg}(f_i) + 1 \lneq \operatorname{ind_{\mathcal{O}}}(f)-1+1$. 
Thus $\operatorname{ind_{\mathcal{O}}}(f_ig_i) \lneq \operatorname{ind_{\mathcal{O}}}(f)$. Also it can be seen that,
$\operatorname{ind_{\mathcal{O}}}(f+g) \leq \operatorname{max} \{\operatorname{ind_{\mathcal{O}}}(f),\operatorname{ind_{\mathcal{O}}}(g)\}$,
 when either $\operatorname{ind_{\mathcal{O}}}(f) \geq 1$ or $\operatorname{ind_{\mathcal{O}}}(g) \geq 1$. Thus,\\
 $\operatorname{ind_{\mathcal{O}}}(\sum_{i=1}^{s} f_ig_i) \lneq \operatorname{ind_{\mathcal{O}}}(f)$. This is a contradiction.
 Hence $\operatorname{max}\{ \operatorname{deg}(f_i)\} = \operatorname{ind_{\mathcal{O}}}(f)-1$.

\noindent (iii) $\Rightarrow$ (iv) Since each  $g_i \in \mathfrak{a}$ and $\operatorname{BF}_\mathcal{O}(g_i) = c_ix^{\beta_i}$, we have $\langle c_1x^{\beta_1}, \ldots, c_sx^{\beta_s} \rangle \subseteq \operatorname{BF}_{\mathcal{O}}(\mathfrak{a})$. 
Consider a polynomial $f \in \mathfrak{a}$. Suppose $\operatorname{ind}_{\mathcal{O}}(f) \geq 1$, then by Definitions~\ref{Index definition} and~\ref{border form definition} each term in $\operatorname{BF}_{\mathcal{O}}(f)$ is divisible by $cx^{\beta} \in \partial\mathcal{O}$. Hence, it follows that $\operatorname{BF}_{\mathcal{O}}(f) \in \langle c_1x^{\beta_1}, \ldots, c_sx^{\beta_s} \rangle$. 

\noindent
Let $\mathcal{I}$ be the proper order function associated with the order ideal $\mathcal{O}$. 
Now let us assume that the there exists a polynomial $f \in \mathfrak{a} \setminus \{0\}$ such that $ \operatorname{ind}_{\mathcal{O}}(f) = 0$ i.e., $f = \sum_{i = 1}^{t}c_ix^{\alpha_i} $ where 
$c_i \notin \mathcal{I}_{x^{\alpha_i}}$. Then by hypothesis, there exist $f_i$s, $1 \leq i \leq s$, such that $f = \sum_{i = 1}^{s} f_ig_i$ and $\operatorname{max}\{\operatorname{deg}(f_i)\ |\ f_ig_i \neq 0, i = 1, \ldots , s  \} = 0 - 1 = -1$. This is not possible since $\operatorname{deg}(f) \geq 0$ for all $f \in \mathbb{A}[x_1, \ldots, x_n]$. This implies that $f = 0$ which is a contradiction. 
Therefore,  $ \operatorname{ind}_{\mathcal{O}}(f) \geq 1$. Thus, $\operatorname{BF}_{\mathcal{O}}(\mathfrak{a}) \subseteq \langle c_1x^{\beta_1}, \ldots, c_sx^{\beta_s} \rangle$. The claim follows.

\noindent (iv) $\Rightarrow$ (i). Consider a polynomial $f \in A[x_1,\ldots,x_n]$. By the border division algorithm, we have $f_1,\ldots,f_s \in 
A[x_1,\ldots,x_n]$ and $a_1,\ldots,a_t \in A$ such that 
\[
 f=f_1g_1+\cdots+f_sg_s+a_1x^{\alpha_1}+\cdots+a_tx^{\alpha_t}.
\]
\noindent Since $\sum_{i=1}^{s} f_{i}g_{i} \in \mathfrak{a}$, $f = h \text{ mod } \mathfrak{a}$, 
where $h=\sum_{j=1}^{t} a_{j}x^{\alpha_{j}}$. 
We are given that the $\mathcal{O}$-border prebasis, $G$ is acyclic. Without loss of generality, let
us assume that $G$ is well ordered. Find the smallest $i$ for which the border term in  $g_i$ belongs to $\partial\mathcal{O}_s$ and
assume that the monomial in the border term is $x^{\alpha_1}$. Let $G_{1} \subseteq G$ 
represent all the polynomials for which the border monomial is $x^{\alpha_1}$ and let $|G_1|=s_1$. 

Let the ideal $\mathcal{I}_{x^{\alpha_1}}$ be generated by $\{ c_1,\ldots,c_{s_1} \}$ and let $b_{1} \in C_{\mathcal{I}_{x^{\alpha_1}}}$ be the coset representative of $a_1 + \mathcal{I}_{x^{\alpha_{1}}}$. 
Then there exist $d_1,\ldots,d_{s_1} \in A$ such that 
\[
 (a_1-b_1) = d_1c_1+\cdots+d_{s_1}c_{s_1}.
\]
Let $h_1=h-d_1g_1+\cdots+d_{s_1}g_{s_1}$.
Therefore we have, 
\[
 h_1=b_1x^{\alpha_1}+b'_{2}x^{\alpha_2} + \cdots + b'_{t}x^{\alpha_t}, 
\]
where $b'_{i} \in A$, $i \in \{2, \ldots, t\}$. Further, $h_1 = h \text{ mod } \mathfrak{a}$. Repeating the above process for the remaining monomials in 
$\partial\mathcal{O}_s$ in the same sequence as the well ordered basis, we get
\[
 h_{s'} = b_1x^{\alpha_1}+\ldots+b_{s'}x^{\alpha_{s'}}+b_{s'+1}x^{\alpha_{s+1}}+\cdots+b_{t}x^{\alpha_t}, 
\]
where 
each $b_i$ is a coset representative in $C_{\mathcal{I}_{x^{\alpha_i}}}$. Note that for 
$x^{\alpha_i}$, ${s'}+1 \leq i \leq t$, $\mathcal{I}_{x^{\alpha_i}} = \{0\}$ and $b_i\in A$. The acyclicity of the basis ensures that 
at each stage, $i$, all $b_j,\ j \lneq i$ will not be modified. Further, at
every stage $i$, the intermediate polynomial, $h_{i} = h_{i-1} \text{ mod } \mathfrak{a}$. Therefore, 
$h_{s'} = h \text{ mod } \mathfrak{a}$ which implies that $f = h_{s'} \text{ mod } \mathfrak{a}$. 
Further, each $b_i$ is a coset representative in $C_{\mathcal{I}_{x^{\alpha_i}}}$ where $x^{\alpha_i} \in \Mon(\mathcal{O})$.
Hence, $A[x_1,\ldots,x_n]/\mathfrak{a} = \{\sum_{i=1}^{t} a_ix^{\alpha_i}+ \mathfrak{a}\ |\ a_i \in C_{\mathcal{I}_{x^{\alpha_i}}}\}$.

\noindent
To prove the second condition of the order span definition (Definition~\ref{weakstarbasis}), consider a polynomial $f=\sum_{i=1}^{t} a_ix^{\alpha_i} \in \mathfrak{a}$. \
Then, there exists an $i \in \{1,\ldots,t\}$ such that
$a_ix^{\alpha_i} \in \operatorname{BF_{\mathcal{O}}}(f)$. By hypothesis we have, $\operatorname{BF_{\mathcal{O}}}(\mathfrak{a})=
\langle c_1x^{\beta_1},\ldots,c_sx^{\beta_s} \rangle$. Since $\operatorname{BF_{\mathcal{O}}}(\mathfrak{a})$ is an ideal 
generated by terms, we have $a_ix^{\alpha_i} \in \operatorname{BF_{\mathcal{O}}}(\mathfrak{a})$. Thus, there exists terms
$d_ix^{\gamma_i}$, $1 \leq i \leq s$ such that
\[
 a_ix^{\alpha_i} = \sum_{i=1}^{s} (d_ix^{\gamma_i})(c_ix^{\beta_i}). 
\]
Since for all $x^{\beta}|x^{\alpha_i}$ $\mathcal{I}_{x^{\beta}} \subseteq \mathcal{I}_{x^{\alpha_{i}}}$, we have $a_i \in \mathcal{I}_{x^{\alpha_{i}}}$.

\noindent Consider a proper order function, $\mathcal{I}^{'}$ such that for any $x^\alpha \in \mathrm{Mon}(A[x_1, \ldots, x_n])$, either 
 $\mathcal{I}^{'}_{x^\alpha} \subsetneq \mathcal{I}_{x^\alpha}$ or $\mathcal{I}^{'}_{x^\alpha} = \mathcal{I}_{x^\alpha}$. Let $\mathcal{O}^{'}$ be the order ideal associated with $\mathcal{I}^{'}$ and 
  $\partial\mathcal{O}^{'}=\{c_{1}^{'}x^{\beta_1},\ldots,c_{l}^{'}x^{\beta_l}\}$ be the corresponding border.
 Assume that $\mathrm{Mon}(\mathcal{O}^{'})$ satisfies (i) and (ii). 
 Consider the ideal, $\mathfrak{A} = \langle c_{1}^{'}x^{\beta_1},\ldots,c_{l}^{'}x^{\beta_l} \rangle$. We have, $\mathfrak{A} \subsetneq \operatorname{BF_{\mathcal{O}}}(\mathfrak{a})$ since
  $\mathcal{I}^{'}_{x^\alpha} \subsetneq \mathcal{I}_{x^\alpha}$ 
  % this was here previously -- $\mathcal{I}_{x^\alpha} \subsetneq \mathcal{I}^{'}_{x^\alpha}$ 
  for some $x^\alpha$. Consider $c_\omega x^\omega \in \operatorname{BF_{\mathcal{O}}}(\mathfrak{a}) \setminus \mathfrak{A}$.
 Let $h$ denote the $\mathrm{rem}_{\mathcal{O}, G}(c_\omega x^\omega)$. Then each non zero term in $h$ is of the form $d_\gamma x^\gamma$ such that $d_\gamma \in A/\mathcal{I}_{x^\gamma}$ and 
 $x^\gamma \in \mathrm{Mon}(\mathcal{O})$. Since $d_\gamma \notin \mathcal{I}_{x^\gamma}$, it is not an element of  $\mathcal{I}^{'}_{x^\gamma}$. Also, $c_\omega \notin \mathcal{I}^{'}_{x^\omega}$. 
 Therefore, $\mathrm{Mon}(\mathcal{O}^{'})$ fails to satisfy Condition (ii) of the definition of order span for $c_\omega x^\omega - h$  and therefore we have a contradiction. 
% To prove that the third condition of Definition~\ref{weakstarbasis} holds we assume that there exists a set $B \subsetneq \Mon(\mathcal{O})$ such that it satisfies
% conditions (i) and (ii) of Definition~\ref{weakstarbasis}. Let $x^\alpha \in \Mon(\mathcal{O}) \setminus B$. 
% Here, we need to consider two cases.
% \begin{enumerate}[(a)]
% \item  ${\mathcal{I}^{(G)}}_{x^{\alpha}}   = \{0\}$. 
% This implies $x^\alpha \notin \partial\mathcal{O}$ and by our hypothesis $x^\alpha \notin \operatorname{BF}_{\mathcal{O}}(\mathfrak{a})$.
% Therefore, $x^\alpha \notin \mathfrak{a}$.
% Since $x^\alpha \notin B$, $\Span_A(B + \mathfrak{a}) \neq A[x_{1},\ldots,x_{n}]/\mathfrak{a}$ and $B$ is not a $\text{weak}^{+}$ basis. 
% \item  ${\mathcal{I}^{(G)}}_{x^{\alpha}}   \neq \{0\}$. 
% %This implies there is atleast one $g_j \in G$ such that $\mathrm{Lm}(g_j) \mid x^{\alpha_i}$.  
% Since ${\mathcal{I}^{(G)}}_{x^{\alpha}}  \neq  \langle 1 \rangle$ there exists some nonzero $b \in C_{\mathcal{I}^{(G)}_{x^{\alpha}}} $ 
% such that $bx^\alpha \notin \partial\mathcal{O}$. This implies that   $x^\alpha \notin \operatorname{BF}_{\mathcal{O}}(\mathfrak{a})$ and
% therefore, $x^\alpha \notin \mathfrak{a}$. 
% %such that 
% %$b \notin \langle \mathrm{Lc}(g_{i_1}), \ldots, \mathrm{Lc}(g_{i_l}) \rangle$, where $\Lm(g_{i_j}) \mid x^\alpha$, $j \in \{1, \ldots, l\}$. 
% %Thus, 
% Since $x^\alpha \notin B$, $\Span_A(B + \mathfrak{a}) \neq A[x_{1},\ldots,x_{n}]/\mathfrak{a}$ and  $B$ is not a $\text{weak}^{+}$ basis. 
% \end{enumerate}
Thus $G=\{g_1,\ldots,g_s\}$ is an $\mathcal{O}$-border basis for $\mathfrak{a}$.
\end{proof}

\section{A Full Example}
\label{Example section}
In this section, we illustrate the concepts given in this paper with an example.
\begin{example}\label{detailed_example}
Let us consider the polynomial ring, $\mathbb{Z}[x,y]$. Let $\mathcal{I}$ be an order function such that $\mathcal{I}_1 = \{0\}$, $\mathcal{I}_x = \{0\}$,
$\mathcal{I}_y = \{0\}$, $\mathcal{I}_{xy} = \{0\}$, $\mathcal{I}_{y^2} = \{0\}$, $\mathcal{I}_{x^2} = \langle 2 \rangle$, $\mathcal{I}_{x^2y} = \langle 2 \rangle$ 
and the rest of the monomials map to $\langle 1 \rangle$. Therefore, $C_{\mathcal{I}_{1}}$ = $C_{\mathcal{I}_{x}}$ = $C_{\mathcal{I}_{y}}$
= $C_{\mathcal{I}_{xy}}$ = $C_{\mathcal{I}_{y^2}}$ = $\mathbb{Z}$ and $C_{\mathcal{I}_{x^2}}$ = $C_{\mathcal{I}_{x^2y}}$ = $\{0,1\}$.
% \[
%  B=\{(\phi,1),(\phi,x),(\phi,y),(\phi,xy),(\phi,y^2),(\langle 2 \rangle, x^2),
% (\langle 2 \rangle, x^2y)\}
% \]
%  be a finite set of order tuples. 
 The set, 
\begin{align*}
 \mathcal{O}=&\{a_1,a_2x,a_3y,a_4xy,a_5y^2,a_6x^2,a_7x^2y\ | \\ &\ a_1,a_2,a_3,a_4,a_5 \in \mathbb{Z},\ a_6,a_7 \in \{0,1\} \}
\end{align*}
is an order ideal corresponding to $\mathcal{I}$. The monomial part of $\mathcal{O}$ is the set $\Mon(\mathcal{O})=\{1,x,y,x^2,y^2,xy,x^2y\}$.
The scalar border of the order ideal is $\partial\mathcal{O}_{s} = \{2x^2,2x^2y\}$ and the monomial border is
$\partial\mathcal{O}_{m}=\{x^3,y^3,xy^2,x^2y^2,x^3y\}$. Thus the border of $\mathcal{O}$ is the union of the scalar border and the
monomial border, i.e.,
\[
 \partial\mathcal{O}=\partial\mathcal{O}_{m} \cup \partial\mathcal{O}_{s}=\{x^3,y^3,xy^2,x^2y^2,x^3y,2x^2,2x^2y\}.
\]
Consider the set $G=\{g_1,g_2,g_3,g_4,g_5,g_6,g_7\}$, where
$g_{1}=x^3-x, 
\ g_{2}=y^3-y,
\ g_{3}=xy^2-xy, 
\ g_{4}=x^2y^2-x^2y, 
\ g_{5}=x^3y-xy, 
\ g_{6}=2x^2y-y^2-y \text{ and }
 g_{7}=2x^2+2xy-y^2-2x-y$.
The set $G$ is an $\mathcal{O}$-border prebasis for the ideal
$\mathfrak{a} = \langle g_1,g_2,g_3,g_4,g_5,g_6,g_7 \rangle$. 
It is also clear that the $\mathcal{O}$-border prebasis satisfies the properties of acyclicity. Hence $G$ is an acyclic 
$\mathcal{O}$-border prebasis. 

\noindent
The set $G$ is a Gr\"obner basis for $\mathfrak{a}$ with deglex order with $x>y$. 
The proof of Theorem~\ref{BBexist} implies that the set $G$ is an acyclic $\mathcal{O}$-border basis for $\mathfrak{a}$. 
Hence the border form ideal of $\mathfrak{a}$ is generated by the border terms i.e.,
\begin{displaymath}
   \operatorname{BF}_{\mathcal{O}}(\mathfrak{a})= \langle x^3,y^3,xy^2,x^2y^2,x^3y,2x^2,2x^2y \rangle.
\end{displaymath}

\noindent
Now we demonstrate the border division algorithm (Algorithm~\ref{BorderDiv}) for a polynomial, $f=x^4+2x^3y^2+x^2+4xy+15$ w.r.t. $G$.
We have $\Mon(\mathcal{O})=\{x^{\alpha_1},\ldots,x^{\alpha_{7}}\}$, where $x^{\alpha_1}=1$,
$x^{\alpha_2}=x$,
$x^{\alpha_3}=y$,
$x^{\alpha_4}=x^2$,
$x^{\alpha_5}=y^2$,
$x^{\alpha_6}=xy$ and 
$x^{\alpha_7}=x^2y$. The monomial border is $\partial\mathcal{O}_{m}=\{b_1,\ldots,b_5\}$ and the scalar border is 
$\partial\mathcal{O}_{s}=\{b_6,b_7\}$, where 
$b_1=x^3$,
$b_2=y^3$,
$b_3=xy^2$,
$b_4=x^2y^2$,
$b_5=x^3y$,
$b_6=2x^2y$ and
$b_7=2xy$.
\begin{enumerate}
\item Initialize $f_1=f_2=f_3=f_4=f_5=f_6=f_7=0$, $l_1=l_2=l_3=l_4=l_5=l_6=l_7=0$ and $h=x^4+2x^3y^2+x^2+4xy+15$.
\item Since $\operatorname{ind}_{\mathcal{O}}(h)=2$, Step 5 of the algorithm is executed. We have  $x^4=xb_1$ and 
$\operatorname{deg}(x) = \operatorname{ind}_{\mathcal{O}}(h)-1$. Hence, we have $f_1=f_1+x$ and $h=x^4+2x^3y^2+x^2+4xy+15-x(x^3-x)$.
Thus, $h=2x^3y^2+2x^2+4xy+15$. Return to Step 2.
\item Again, $\operatorname{ind}_{\mathcal{O}}(h)=2$ and we return to Step 5 of the algorithm. We have  $x^3y^2=xb_4$ and 
$\operatorname{deg}(x) = \operatorname{ind}_{\mathcal{O}}(h)-1$. After the reduction step we have $f_4=f_4+2x$ and 
$h=2x^3y^2+2x^2+4xy+15-2x(x^2y^2-x^2y)=2x^3y+2x^2+4xy+15$. We return to Step 2.
\item In this step, $\operatorname{ind}_{\mathcal{O}}(h)=1$ and the polynomial $h$ has the monomial $x^3y$ in its support. 
Since $x^3y \in \partial\mathcal{O}_{m}$, we go to Step 5. We have $x^3y=1\cdot b_5$ and $\operatorname{deg}(1) = 
\operatorname{ind}_{\mathcal{O}}(h)-1$. We perform the operation $f_5=f_5+2$ and $h=2x^3y+2x^2+4xy+15-2(x^3y-xy)$. 
Hence, $h=2x^2+6xy+15$. We return to Step 2.
\item We have $\operatorname{ind}_{\mathcal{O}}(h)=1$ and none of the terms in $h$ are in the monomial border,
$\partial\mathcal{O}_{m}$. Hence, we perform Step 4 of the algorithm. We have $2x^2=1\cdot b_1$. Thus, we have $f_7=f_7+1$ and
$h=2x^2+6xy+15-1(2x^2+2xy-y^2-2x-y)$. Hence, $h=4xy+y^2+2x+y+15$.
\item We have $\operatorname{ind}_{\mathcal{O}}(h)=0$ and Step 3 of the algorithm is executed. We have
$l_1=l_1+15$,
$l_2=l_2+2$,
$l_3=l_3+1$,
$l_4=l_4+0$,
$l_5=l_5+1$,
$l_6=l_6+4$ and
$l_7=l_7+0$. The algorithm terminates and returns $(f_1,\ldots,f_7,l_1,\ldots,l_7)$.
\end{enumerate}
Thus we have the following representation for $f$,
\begin{align*}
 f=&xg_1+0g_2+0g_3+1g_4+2g_5+0g_6+1g_7+ \\ &15+2x+y+0x^2+1y^2+4xy+0x^2y.
\end{align*}
The $\mathcal{O}$-remainder of $f$ is $\operatorname{rem}_{\mathcal{O},G}(f)=15+2x+y+y^2+4xy$. Since the remainder is
not equal to zero, by Proposition~\ref{IdealRem}, $f \notin \mathfrak{a}$. For this example, the normal form of $f$ is 
equal to the $\mathcal{O}$-remainder,
\[
 \operatorname{NF}_{\mathcal{O},G}(f)=\operatorname{rem}_{\mathcal{O},G}(f)=15+2x+y+y^2+4xy.
\]
\end{example}

%========================================================
\section{Concluding Remarks}
\label{Concluding remarks}
\noindent
The theory of border bases is popular in the symbolic computation
community owing to its numerical stability and mathematical
elegance. In~\citep{FrancisDukkipati:2014:OnReducedGrobnerBasisAndMBtheorem}
it has been shown that border bases can be extended to polynomial
rings over rings if the corresponding residue class ring is free and if
one can find its free representation with respect to a monomial order by
using Gr\"obner bases. In this paper we attempted to extend the theory
to a much more general case.

%In this paper, we introduced the notion of coefficient ideal mapping
%and used that to extend the theory of border bases to polynomial
%rings over Noetherian rings.

%There are two cases here.
%When the residue class ring is free, the theorem 
% gives an algorithm to determine the $A$-module basis and when it is
% not free it gives a generating set  that satisfies a weak form of
% linear independence property with respect to a coefficent ideal
% mapping. 
%  This generating set gives an isomorphic structure for the residue
%  class polynomial ring over $A$ as a Cartesian product of residue
%  class rings in $A$.
% \new comments
%This generalization plays a crucial role in extending the concept of
%border bases to polynomial rings over Noetherian rings.
%new comments stop here
A theory was built for a special class of
generators of the ideal called acyclic border bases. We proved that
the set of acyclic border bases contains all the reduced Gr\"obner
bases associated with all term orders. 
 %which shows that considering this set is not a serious limitation. 
 %It would be an interesting problem  to prove/disprove whether the set of reduced Gr\"obner bases
% is a strict subset of this class.
A future direction in this work is to determine an algorithmic
characterization for border bases in this case.
%\begin{itemize}
% \item We have not given an algorithmic characterization for border bases  for an ideal $\mathfrak{a}$ in $A[x_1,\ldots,x_n]$.
% \item Acyclic border basis contains all the reduced Gr\"obner basis associated with all term orders. Since we do not have an algorithm for characterization of border
% bases, we cannot establish if the set of acyclic border bases is bigger. It would be an interesting problem to prove/disprove that the set of reduced Gr\"obner basis
% is a strict subset of acyclic brder basis.
% \item Examples where $A$ is $\Bbbk[x]$ and $\Bbbk[x_{1},\ldots,x_{n}]$
%\end{itemize} 
%=========================================================

{\footnotesize
\bibliographystyle{jtbnew}
\bibliography{grobner}
}

\begin{example}\label{border_prebases_example}
We consider Example~\ref{exampleref}. 
%We have the order ideal $\mathcal{O}=\{a_1,a_2x,a_3y,a_4x^{2}$ $|\ a_1 \in \mathbb{Z},\ a_2 \in \mathbb{Z}_{4},\ a_3 \in \mathbb{Z}_{3},\  a_4 \in \mathbb{Z}_{2}\}$ and the $\mathcal{O}$-border
%$\partial\mathcal{O}=\{xy,y^{2},x^{3},x^{2}y,4x,3y,2x^{2}\}$. 
The set $G=\{g_1,\ldots,g_7\}$, where $g_1=xy-x$, $g_2=y^2-y$, $g_3=x^3-2y$, $g_4=x^2y-x^2+10$,  $g_5=4x-2y$, $g_6=3y-3x$ and $g_7=2x^2-x+5$ is an 
$\mathcal{O}$-border prebasis but it is not acyclic. 
Let $G'=\{g'_1,\ldots,g'_7\}$ 
where $g'_1=xy-x$, $g'_2=y^2-y$, $g'_3=x^3-x^2+6$, $g'_4=x^2y-y+5$, $g'_5=4x-7$, $g'_6=3y-x$ and $g'_7=2x^2-2y-3x$ which is also a $\mathcal{O}$-border prebasis but it is acyclic since the permutation of $G'$, $\{g'_1,g'_2,g'_3,g'_4,g'_7,g'_6,g'_5\}$ satisfies  
the acyclicity condition.
\end{example}

\begin{example}\label{border_prebases_example2}
Consider Example~\ref{exampleref2}. 
%We have the order ideal $\mathcal{O'} = \{a_1,\ a_2x,\ a_3y,\ a_4z,$ $a_5x^2,\ a_6xy,\ a_7xz \  |\ 
%a_1 \in \Bbbk[u_1,u_2],\ a_2 \in \Bbbk[u_1,u_2],\ a_3 \in \Bbbk[u_1,u_2]/\langle u_{1}^{2} \rangle,
 %\ a_4 \in \Bbbk[u_1,u_2]/\langle u_{2}^{2} - 3u_1 \rangle,\ a_5 \in \Bbbk[u_1,u_2],\ a_6 \in \Bbbk[u_1,u_2]/\langle u_{1}^{2} , u_{2}^{2}-1 \rangle,\ 
 % a_7 \in \Bbbk[u_1,u_2]/\langle u_{1} , u_2  \rangle \}$ and the $\mathcal{O'}$-border, 
 %$\partial \mathcal{O'} = \{ x^3,x^2y,x^2z,xy^2,xyz,xz^2,y^2,yz,z^2,u_{1}^{2}y,(u_{2}^{2}-3u_1)z,u_{1}^{2}xy,(u_{2}^{2}-1)xy,u_1xz,u_2xz \}$. 
The set  $G =\\ \{g_1,\ldots,g_{15}\}$, where $g_1=x^3-3$, $g_2=x^2y-3u_1y$, $g_3=x^2z-2z$, $g_4=xy^2-x+10$,  $g_5=xyz-11xy$, $g_6=xz^2-u_{2}u_{1}^{2}x^2$, $g_7=y^2-x+u_1u_2$, $g_8=yz-3y+2$, 
$g_9=z^2+5xz+11u_1x$, $g_{10}=u_{1}^{2}y+u_2x+3$, $g_{11}=(u_{2}^{2}-3u_1)z-u_{2}^{2}y$, $g_{12}=u_{1}^{2}xy+3u_1x-2z$, $g_{13}=(u_{2}^{2}-1)xy+2x^2$, $g_{14}=u_1xz+3u_1x^2$, 
$g_{15}=u_2xz+2u_{1}xy+4x^2-4z-10u_1y+14$ 
is an acyclic $\mathcal{O}$-border prebasis since the following permutation of $G$, $\{g_1,g_2,g_3,g_4,g_5,g_6,g_7,g_8,g_9,g_{14},g_{15},g_{13},g_{12},g_{11},g_{10}\}$ satisfies  
the acyclicity condition.
\end{example}
\end{document}